\documentclass{elsart1p}
\usepackage{amstext,amsmath,amsfonts,amssymb,amsbsy}
\usepackage{bm}
\usepackage{color,lscape}
\usepackage{graphicx}
\usepackage{multirow}
\usepackage{rotating}

\def\rmDVCS{\mathrm{DVCS}}
\def\rmI{\mathrm{I}}

\def\rmunp{\mathrm{unp}}

\def\CULPM{\sigma_{\mathrm{U{\stackrel{\scriptscriptstyle 
\Leftarrow}{\scriptscriptstyle \Rightarrow}}}}}
\def\CLCPM{\sigma_{\mathrm{{\stackrel{0}{\scriptscriptstyle \leftarrow}}}
{\stackrel{\Leftarrow}{\scriptscriptstyle \Rightarrow}}}}
\def\Lumi{\mathcal{L}\,}

\def\tt{\bm{\eta}}

\def\AC{A_\mathrm{C}}

\def\ALUI{A_{\mathrm{LU},\rmI}}
\def\CalAC{\mathcal{A}_\mathrm{C}}

\def\CalALUDVCS{\mathcal{A}_{\mathrm{LU}}^{\rmDVCS}}
\def\CalALUI{\mathcal{A}_{\mathrm{LU}}^{\rmI}}

\def\ALPM{A_\mathrm{\scriptscriptstyle 
L{\stackrel{\Leftarrow}{\scriptscriptstyle \Rightarrow}}}}
\def\AUL{A_\mathrm{UL}}
\def\ALL{A_\mathrm{LL}}
\def\CalALPM{\mathcal{A}_\mathrm{\scriptscriptstyle 
L{\stackrel{\Leftarrow}{\scriptscriptstyle \Rightarrow}}}}
\def\CalAUL{\mathcal{A}_\mathrm{UL}}
\def\CalALL{\mathcal{A}_\mathrm{LL}}

\def\ALPMcoh{A_{\mathrm{\scriptscriptstyle 
L{\stackrel{\Leftarrow}{\scriptscriptstyle \Rightarrow}}},\mathrm{coh}}}
\def\ALUcoh{A_{\mathrm{LU},\rmI,\mathrm{coh}}}

\def\AP{A_\mathrm{{\stackrel{0}{\scriptscriptstyle 
\leftarrow}}\scriptscriptstyle L}}
\def\ACP{A_\mathrm{{\stackrel{C}{\scriptscriptstyle 
\leftarrow}}\scriptscriptstyle L}}

\def\ALzz{A_\mathrm{Lzz}}
\def\CalALzz{\mathcal{A}_\mathrm{Lzz}}

\def\ACPMM{A_\mathrm{{\stackrel{C}{\scriptscriptstyle \leftarrow}}
{\stackrel{\Leftarrow}{\scriptscriptstyle \Rightarrow}}}}

\def\CalACPM{\mathcal{A}_\mathrm{{\stackrel{C}{\scriptscriptstyle 
\leftarrow}}{\stackrel{\Leftarrow}{\scriptscriptstyle \Rightarrow}}}}
\def\CalAP{\mathcal{A}_\mathrm{{\stackrel{0}{\scriptscriptstyle 
\leftarrow}}\scriptscriptstyle L}}
\def\CalACP{\mathcal{A}_\mathrm{{\stackrel{C}{\scriptscriptstyle 
\leftarrow}}\scriptscriptstyle L}}

\def\CalACPMN{\mathcal{A}_\mathrm{C{\stackrel{\scriptscriptstyle 
\Leftarrow}{\scriptscriptstyle \Rightarrow}}}}

\def\intN{{\mathcal N}}

\def\chk{\textcolor{red}}

\def\amp2{{\cal T}}

\newcommand{\rd}{\mathrm{d}}

\journal{Nuclear Physics B}

\begin{document}

\begin{frontmatter}

\title{Measurement of azimuthal asymmetries associated with deeply virtual 
Compton scattering on a longitudinally polarized deuterium target}

\collab{The HERMES Collaboration}

\author[12,15]{A.~Airapetian}
\author[26]{N.~Akopov}
\author[5]{Z.~Akopov}
\author[6]{E.C.~Aschenauer\thanksref{27}}
\author[25]{W.~Augustyniak}
\author[26]{R.~Avakian}
\author[26]{A.~Avetissian}
\author[5]{E.~Avetisyan}
\author[18]{S.~Belostotski}
\author[10]{N.~Bianchi}

% ...
\address[5]{DESY, 22603 Hamburg, Germany}
\address[6]{DESY, 15738 Zeuthen, Germany}
% ...
\address[10]{Istituto Nazionale di Fisica Nucleare, Laboratori Nazionali di Frascati, 00044 Frascati, Italy}
% ...
\address[12]{Physikalisches Institut, Universit\"at Gie{\ss}en, 35392 Gie{\ss}en, Germany}
% ...
\address[15]{Randall Laboratory of Physics, University of Michigan, Ann Arbor, MI 48109-1040, USA}
% ...
\address[18]{Petersburg Nuclear Physics Institute, Gatchina, Leningrad region 188300, Russia}
% ...
\address[25]{Andrzej Soltan Institute for Nuclear Studies, 00-689 Warsaw, Poland}
\address[26]{Yerevan Physics Institute, 375036 Yerevan, Armenia}

\begin{abstract}
Azimuthal asymmetries in exclusive electroproduction of a real photon from 
a longitudinally polarized deuterium target are measured with respect to 
target polarization alone and with respect to target polarization combined 
with beam helicity and/or beam charge. The asymmetries appear in the 
distribution of the real photons in the azimuthal angle $\phi$ around the 
virtual photon direction, relative to the lepton scattering plane. The 
asymmetries arise from the deeply virtual Compton scattering process and 
its interference with the Bethe-Heitler process. The results for the 
beam-charge and beam-helicity asymmetries from a tensor polarized 
deuterium target with vanishing vector polarization are shown to be 
compatible with those from an unpolarized deuterium target, which is 
expected for incoherent scattering dominant at larger momentum transfer. 
Furthermore, the results for the single target-spin asymmetry and for the 
double-spin asymmetry are found to be compatible with the corresponding 
asymmetries previously measured on a hydrogen target. For coherent 
scattering on the deuteron at small momentum transfer to the target, these 
findings imply that the tensor contribution to the cross section is small. 
Furthermore, the tensor asymmetry is found to be compatible with zero.
\end{abstract}

\begin{keyword}
DIS \sep HERMES experiments \sep GPDs \sep DVCS 
\sep polarized deuterium target
\PACS 13.60.-r \sep 24.85.+p \sep 13.60.Fz \sep 14.20.Dh
\end{keyword}

\end{frontmatter}

\section{Introduction}
\label{sec:Introduction}
Generalized Parton Distributions (GPDs) provide a framework for describing 
the multidimensional structure of the nucleon~\cite{Mul94,Rad97,Ji97}. 
GPDs encompass parton distribution functions and elastic nucleon form 
factors as limiting cases and moments, respectively. Parton distribution 
functions are distributions in longitudinal momentum fraction of partons 
in the nucleon, and are extracted from measurements of inclusive and 
semi-inclusive deep-inelastic scattering. Form factors are related to the 
transverse spatial distribution of charge and magnetization in the 
nucleon. Both form factors and (transverse-momentum-integrated) parton 
distribution functions represent one-dimensional distributions, whereas 
GPDs provide correlated information on transverse spatial and longitudinal 
momentum distributions of partons~\cite{GPD2,Bur00,GPD3,GPD4,GPD5,GPD6}. 
Furthermore, access to the total parton angular momentum contribution to 
the nucleon spin may be provided by GPDs through the Ji 
relation~\cite{Ji97}.

Hard exclusive leptoproduction of a meson or photon, with only an intact 
nucleon or nucleus remaining in the final state, can be described in terms 
of GPDs. GPDs depend on four kinematic variables: $t, x, \xi$, and $Q^2$. 
In this case, $t$ is the Mandelstam variable, or the squared four-momentum 
transfer to the target, given by $t=(p-p^\prime)^2$, where $p$ 
($p^\prime$) is the initial (final) four-momentum of the target. In 
the `infinite' target-momentum frame, $x$ and $\xi$ are related to the 
longitudinal momentum of the parton involved in the interaction as a 
fraction of the target momentum. The variable $x$ is the average momentum 
fraction and the variable $\xi$, known as the skewness, is half the 
difference between the initial and final momentum fractions carried by the 
parton. The evolution of GPDs with $Q^2 \equiv -q^2$, with $q = k - 
k^\prime$ the difference between the four-momenta of the incident and 
scattered leptons, can be calculated in the context of perturbative 
quantum chromodynamics as in the case of parton distribution functions. 
This evolution has been evaluated to leading 
order~\cite{Mul94,Rad97,Ji97,Blu97} and next-to-leading 
order~\cite{BelMul99,BelFreu00,BelMul00} in the strong coupling constant 
$\alpha_s$. The skewness $\xi$ can be related to the Bjorken scaling 
variable $x_B \equiv Q^2/(2p \cdot q)$ through $\xi \simeq x_B/(2-x_B)$ in 
the generalized Bjorken limit of large $Q^2$, and fixed $x_B$ and $t$. 
There is currently no consensus as to how to define $\xi$ in terms of 
experimental observables; hence the experimental results are typically 
reported as projections in $x_B$. The entire $x$ dependences of GPDs are 
generally not experimentally accessible, an exception being the trajectory 
$x = \xi$~\cite{Anikin_Teryaev,Kumericki_Muller}.

GPDs can be constrained through measurements of cross sections and 
asymmetries in exclusive processes such as exclusive photon or meson 
production. In this paper, the Deeply Virtual Compton Scattering (DVCS) 
process, i.e., the hard exclusive production of a real photon, is 
investigated using a longitudinally polarized deuterium target.

The spin-1/2 nucleon is described by four leading-twist 
quark-chirality-conserving GPDs $H$, $E$, $\widetilde{H}$ and 
$\widetilde{E}$~\cite{Mul94,Rad97,Ji97,DVCS2}. In contrast, DVCS leaving 
the spin-1 deuteron intact requires nine GPDs: $H_1$, $H_2$, $H_3$, $H_4$, 
$H_5$, $\widetilde{H}_1$, $\widetilde{H}_2$, $\widetilde{H}_3$ and 
$\widetilde{H}_4$~\cite{berger,theor_deu,theor_bsa1}. In the forward limit 
of vanishing four-momentum transfer to the target nucleon ($t \rightarrow 
0$ and $\xi \rightarrow 0$), the pairs of GPDs ($H$, $H_1$) and 
($\widetilde{H}$, $\widetilde{H}_1$) reduce respectively to quark number 
density and helicity distributions. In this limit the GPD $H_5$, sensitive 
to tensor effects in the deuteron, reduces to the tensor structure 
function $b_1$, which was measured in inclusive deep-inelastic scattering 
on a tensor polarized deuterium target~\cite{hermes:b1}. Both $H_3$ and 
$H_5$ are associated with the 5$\%$ $D$-wave component of the deuteron 
wave function in terms of nucleons~\cite{Lacombe:1981}. In addition to GPD 
$H_1$, they both contribute to the beam-helicity and beam-charge 
asymmetries. The term with GPD $H_5$ dominates in the 
beam-helicity$\otimes$tensor asymmetry in DVCS from a longitudinally 
polarized deuterium target at very small values of $t$~\cite{theor_deu}. 
At this kinematic condition, the asymmetry with respect to target 
polarization is dominated by the term with GPD $\widetilde{H}_1$. Thus, 
the measurement of certain asymmetries in DVCS on a polarized deuterium 
target may provide new constraints for these GPDs.

This paper reports the first observation of azimuthal asymmetries with 
respect to target polarization alone and with respect to target 
polarization combined with beam helicity and/or beam charge, for exclusive 
electroproduction of real photons from a longitudinally polarized 
deuterium target. The asymmetries arise from the DVCS process where the 
photon is radiated by the struck quark, and its interference with the  
Bethe--Heitler (BH) process where the photon is radiated by the initial or 
final state lepton. The resulting asymmetries combine contributions from 
the coherent process $e \, d \to e \, d \, \gamma$, and the incoherent 
process $e \, d \to e \, p \, n \, \gamma$ where in addition a nucleon may 
be excited to a resonance. The coherent reaction contributes mainly at 
very small values of $t$, while the incoherent process dominates 
elsewhere. It is natural to model the incoherent process as scattering on 
only one nucleon in the deuteron, while the other nucleon acts as a 
spectator. Monte Carlo simulations in HERMES kinematic 
conditions~\cite{Bernie} suggest that the proton contributes about 75$\%$ 
of the incoherent yield and the neutron about 25$\%$, and included in 
these, nucleon resonance production contributes about 22$\%$ of the 
incoherent yield. The incoherent reaction on a proton dominates that on a 
neutron because of the suppression of the BH amplitude on the neutron by 
the small elastic electric form factor at low and moderate values of the 
momentum transfer to the target. The dependence of the measured 
asymmetries on the kinematic conditions of the reaction is also presented 
and these results on the deuteron are compared where appropriate with the 
corresponding results obtained on a longitudinally polarized hydrogen 
target~\cite{proton_pol_draft}.

\section{Deeply virtual Compton scattering}
\label{sec:GPDsAndDVCS}

\subsection{Scattering amplitudes}
\label{Amplitudes}
The DVCS process is currently the simplest experimentally accessible 
process that can be used to constrain GPDs. The initial and final states 
of DVCS are indistinguishable from those of the competing BH process. For 
a target of atomic mass number $A$ and no target polarization component 
transverse to the direction of the virtual photon, the general expression 
for the cross section of the coherent reaction $e \, A \to e \, A \, 
\gamma$ or incoherent reaction $e \, A \to e \, (A-1)N \, \gamma$ 
reads~\cite{theor_deu,DVCS0}
\begin{equation} \label {total_gamma_xsect}
\frac{\rd \sigma}{\rd x_{A} \, \rd Q^2 \, \rd |t| \, \rd \phi} =
\frac {x_{A} \, e^6} {32 \, (2 \pi)^4 \, Q^4}
\frac {\left| \amp2 \right|^2} {\sqrt{1 + \varepsilon^2}} \,.
\end{equation}
Here, $x_A \equiv Q^2/(2M_A\nu)$ is the nuclear Bjorken variable, where 
$M_A$ is the mass of the nucleus and $\nu \equiv p \cdot q/M_A$, $e$ is 
the elementary charge, $\varepsilon\equiv 2 x_A M_A/\sqrt{Q^2}$ and 
$\amp2$ is the reaction amplitude. The azimuthal angle of the real photon 
around the virtual-photon direction, relative to the lepton scattering 
plane, is denoted by $\phi$. The cross section contains the coherent 
superposition of BH and DVCS amplitudes:
\begin{equation} \label {eqn:tau}
\left| \amp2 \right|^2 =  \left| \amp2_{\rm BH} +  \amp2_{\rm DVCS} 
\right|^2 = \left| \amp2_{\rm BH} \right|^2 +
\left| \amp2_{\rm DVCS} \right|^2 + \underbrace{
\amp2_{\rm DVCS} \, \amp2_{\rm BH}^* + \amp2_{\rm DVCS}^* \, 
\amp2_{\rm BH}}_{\rm I}\,,
\end{equation}
where `$\rm I$' denotes the BH-DVCS interference term. The BH amplitude is
calculable to leading order in QED using the form factors measured in 
elastic scattering.

The interference term $\rm I$ and the squared DVCS amplitude $\left| 
\amp2_{\rm DVCS} \right|^2$ in Eq.~\ref{eqn:tau} provide experimental 
access to the (complex) DVCS amplitude through measurements of various cross 
section asymmetries as functions of $\phi$~\cite{theor_deu}. Each of the 
three terms of Eq.~\ref{eqn:tau} can be written as a Fourier series in 
$\phi$. In the case that the beam and the target may be longitudinally 
polarized, these terms read
\begin{eqnarray}
&&|\amp2^{}_{\rm BH}|^2 = 
\frac{K_{\rm BH}}{{\cal P}_1(\phi){\cal P}_2(\phi)} \times
\sum_{n=0}^2 c_{n}^{\rm BH} \cos(n\phi) \label{eq:moments-BH}\, , \\
&&|\amp2^{}_{\rm DVCS}|^2 = K_{\rm DVCS} \times \left\{ 
\sum_{n=0}^2 c_{n}^{\rm DVCS} \cos(n\phi) +
\sum_{n=1}^2 s_{n}^{\rm DVCS} \sin (n\phi) \label{eq:moments-DVCS} 
\right\}\, , \\
&&{\rm I} = -\frac{e_\ell K_{\rm I}}{{\cal P}_1(\phi){\cal P}_2(\phi)} 
\times \left\{\sum_{n=0}^3 c_{n}^{\rm I} \cos(n\phi) +
\sum_{n=1}^3 s_{n}^{\rm I} \sin(n\phi) \right\}\,. \label{eq:moments-I}
\end{eqnarray}
The symbols $K_{\rm BH} = \frac{1}{x_{A}^2 \, t \, {(1 + 
\varepsilon^2)}^2}$, $K_{\rm DVCS} = \frac{1}{Q^2}$ and $K_{\rm I} = 
\frac{1}{x_{A} \, y \, t}$ denote kinematic factors, where $y \equiv p 
\cdot q/(p \cdot k)$, and $e_\ell$ stands for the (signed) lepton charge 
in units of the elementary charge. In the case of unpolarized beam and 
target, certain coefficients vanish. All Fourier coefficients $c_n$ and 
$s_n$ in Eqs.~\ref{eq:moments-BH}--\ref{eq:moments-I} depend on the 
longitudinal target polarization, with some also having a dependence on 
the beam helicity. The coefficients $c_{n}^{\rm BH}$ in 
Eq.~\ref{eq:moments-BH} depend on electromagnetic form factors of the 
target, while the DVCS (interference) coefficients $c_{n}^{\rm DVCS}$ 
($c_{n}^{\rm I}$) and $s_{n}^{\rm DVCS}$ ($s_{n}^{\rm I}$) involve various 
GPDs. The squared BH and interference terms in Eqs.~\ref{eq:moments-BH} 
and \ref{eq:moments-I} have an additional $\phi$ dependence in the 
denominator due to the lepton propagators ${\cal P}_{1}(\phi)$ and ${\cal 
P}_{2}(\phi)$~\cite{DVCS0,DVCS2}. The Fourier coefficients $c_n^{\rm I}$ 
and $s_n^{\rm I}$ in Eq.~\ref{eq:moments-I} can be expressed as linear 
combinations of Compton Form Factors (CFFs)~\cite{theor_deu}, while the 
coefficients $c_{n}^{\rm DVCS}$ and $s_{n}^{\rm DVCS}$ are bilinear in the 
CFFs. Such CFFs are convolutions of the corresponding GPDs with the hard 
scattering coefficient functions.

For a longitudinally (L) polarized lepton beam scattered from an 
unpolarized target, the beam-charge asymmetry $\CalAC$ and the 
{\it charge-difference} beam-helicity asymmetry $\CalALUI$ (sensitive to 
the interference term) and {\it charge-average} beam-helicity asymmetry 
$\CalALUDVCS$ (sensitive to the squared DVCS term) can be measured if all 
four combinations of beam charge and helicity are 
available~\cite{proton_unpol_draft}. Results of their Fourier amplitudes 
for an unpolarized deuterium target were recently published by 
HERMES~\cite{deuteron_unpol_draft}.

Unfortunately, the present data set for a longitudinally polarized target 
does not include all four combinations of beam charge and sign of beam 
polarization. Therefore, the beam-helicity asymmetries presented 
in this paper are {\it single-charge} observables, which entangle the 
interference and squared DVCS term. Fortunately, measurements of {\it 
charge-averaged} beam-helicity asymmetries on 
hydrogen~\cite{proton_unpol_draft} and 
deuterium~\cite{deuteron_unpol_draft} targets showed that the contribution 
by the squared DVCS term is negligible in HERMES kinematic conditions, at 
the precision of these measurements.

\subsection{DVCS on the deuteron}
\label{dvcs_deutron}
For coherent scattering on a spin-1 target nucleus polarized 
longitudinally with respect to the virtual photon direction, and with spin 
projection $\Lambda = \pm 1,0$, the following decomposition of the Fourier 
coefficients appearing in Eqs.~\ref{eq:moments-BH}--\ref{eq:moments-I} is 
introduced~\cite{theor_deu}:
\begin{equation}
c_{n}^{\rm R}(\Lambda)=\frac{3}{2}\Lambda^{2}c_{n,\rmunp}^{\rm R}+\Lambda 
c_{n,\rm LP}^{\rm R} + (1-\frac{3}{2}\Lambda^{2})c_{n,\rm LLP}^{\rm R}
\label{eq:coeff_decomp}
\end{equation}
with $\rm R \in \{\rm BH,\rm DVCS,\rm I\}$, and similarly for the $s_n$ 
coefficients with $\rm R \in \{\rm DVCS,\rm I\}$. The subscript `unp' 
denotes unpolarized and `LP' and `LLP' denote respectively vector and 
tensor terms for parts of the cross section related to longitudinal 
polarization. For an unpolarized target nucleus, one recovers the value 
$[c^{\rm R}_n(\Lambda = -1)+c^{\rm R}_n(\Lambda = 0)+c^{\rm R}_n(\Lambda 
= +1)]/3=c^{\rm R}_{n,\rmunp}$. A purely tensor-polarized target nucleus 
with $\Lambda = 0$ results in $c^{\rm R}_{n}=c^{\rm R}_{n, \rm {LLP}}$, 
while for $\Lambda \ne 0$ all coefficients contribute.

Equation~\ref{eq:coeff_decomp} is applicable only for purely polarized 
states with $\Lambda = \pm 1,0$. In a real experiment, the longitudinally 
polarized deuterium target contains a mixture of these pure polarized 
states, characterized by vector and tensor polarizations $P_z$ and 
$P_{zz}$ defined as
\begin{equation}
P_{z} = \frac{n^+ - n^-}{n^+ + n^- + n^0}\, , \,\,\, 
P_{zz} = \frac{n^+ + n^- - 2n^0}{n^+ + n^- + n^0}\, ,
\label{eq:pz_pzz}
\end{equation}
where $n^+$, $n^-$ and $n^0$ are the populations of the state with 
$\Lambda$ = +1, $-1$ and 0, respectively.

For a lepton beam with given longitudinal beam polarization $P_\ell$ 
scattering coherently on a deuterium target with given vector and tensor 
polarizations $P_{z}$ and $P_{zz}$, the Fourier series of the squared 
reaction amplitude reads, using the spin decompositions of 
Eq.~\ref{eq:coeff_decomp},
\begin{align}
\label{eq:moments-BHVT}
& \begin{aligned}[b] |\amp2^{}_{\rm BH}|^2 &= 
\frac{K_{\rm BH}}{{\cal P}_1(\phi){\cal P}_2(\phi)}
\Bigg\{\sum_{n=0}^2 c_{n,\rmunp}^{\rm BH} \cos(n\phi) \\
& + \, P_z P_\ell \sum_{n=0}^1 c_{n,\rm LP}^{\rm BH} \cos(n\phi)
 + \, \frac{1}{2} \, P_{zz} \sum_{n=0}^2 (c_{n,\rmunp}^{\rm 
BH} - c_{n,\rm LLP}^{\rm BH}) \cos(n\phi) \Bigg\}\, , \end{aligned} \\
\label{eq:moments-DVT} 
& \begin{aligned}[b] |\amp2^{}_{\rm DVCS}|^2 &= K_{\rm DVCS}
\Bigg\{\sum_{n=0}^2 c_{n,\rmunp}^{\rm DVCS} \cos(n\phi) + P_\ell 
\, s_{1,\rmunp}^{\rm DVCS} \sin \phi \\
& + \, P_z\bigg[P_\ell \sum_{n=0}^1 c_{n,\rm LP}^{\rm DVCS} \cos(n\phi) +
\sum_{n=1}^2 s_{n,\rm LP}^{\rm DVCS} \sin(n\phi)\bigg] \\
& + \, \frac{1}{2} \, P_{zz} \bigg[\sum_{n=0}^2 (c_{n,\rmunp}^{\rm DVCS} 
- c_{n,\rm LLP}^{\rm DVCS}) \cos(n\phi) + P_\ell 
\, (s_{1,\rmunp}^{\rm DVCS} -
s_{1,\rm LLP}^{\rm DVCS}) \sin \phi\bigg]\Bigg\}\, , \end{aligned} \\
\label{eq:moments-IVT}
& \begin{aligned}[b] {\rm I} &= -\frac{e_\ell K_{\rm I}}{{\cal 
P}_1(\phi){\cal P}_2(\phi)}
\Bigg\{\sum_{n=0}^3 c_{n,\rmunp}^{\rm I} \cos(n\phi) + P_\ell 
\sum_{n=1}^2 s_{n,\rmunp}^{\rm I} \sin(n\phi) \\
& + \, P_z\bigg[P_\ell \sum_{n=0}^2 c_{n,\rm LP}^{\rm I} \cos(n\phi) +
\sum_{n=1}^3 s_{n,\rm LP}^{\rm I} \sin(n\phi)\bigg] \\
& + \, \frac{1}{2} \, P_{zz} \bigg[\sum_{n=0}^3 (c_{n,\rmunp}^{\rm I} -
c_{n,\rm LLP}^{\rm I}) \cos(n\phi) + P_\ell \sum_{n=1}^2 
(s_{n,\rmunp}^{\rm I} - s_{n,\rm LLP}^{\rm I}) \sin(n\phi)\bigg]\Bigg\} 
\,. \end{aligned}
\end{align}
Note that the beam polarization $P_\ell$ and the target vector and tensor 
polarizations $P_{z}$ and $P_{zz}$ are here factored out of the 
corresponding Fourier coefficients in 
Eqs.~\ref{eq:moments-BH}--\ref{eq:moments-I}, thus leaving only the 
dynamical kinematic dependences encoded in the Fourier coefficients in 
Eqs.~\ref{eq:moments-BHVT}--\ref{eq:moments-IVT}.

\subsection{Asymmetries on the deuteron}
\label{Assymetries}
For data with longitudinal polarization of both beam and target, the 
following notation is introduced: $\rightarrow$ ($\leftarrow$) to denote 
positive (negative) beam helicity, and $\Rightarrow$ and $\Leftarrow$ to 
denote the deuteron target vector-polarization direction anti-parallel and 
parallel to the beam momentum direction in the target rest frame. In 
contrast to lepton scattering off longitudinally polarized 
hydrogen~\cite{proton_pol_draft}, there are many more observables 
(asymmetries) in the case of deuterium. They may be classified according 
to whether the cross section for $\Lambda = 0$ explicitly appears in the 
definition of this asymmetry. An example of the `incomplete' asymmetries 
where it does not appear is the beam-helicity asymmetry 
$\CalALPM(e_{\ell},P_{zz},\phi)$, defined for beam charge $e_\ell$ and 
tensor polarization $P_{zz}$ as

\begin{multline}
\CalALPM(e_\ell,P_{zz},\phi) \equiv \\
\frac
{\left[\rd\sigma^{\stackrel{\rightarrow}{\Rightarrow}}(e_\ell,P_{zz},\phi) +
\rd\sigma^{\stackrel{\rightarrow}{\Leftarrow}}(e_\ell,P_{zz},\phi)\right]
-\left[\rd\sigma^{\stackrel{\leftarrow}{\Rightarrow}}(e_\ell,P_{zz},\phi) +
\rd\sigma^{\stackrel{\leftarrow}{\Leftarrow}}(e_\ell,P_{zz},\phi)\right]}
{\left[\rd\sigma^{\stackrel{\rightarrow}{\Rightarrow}}(e_\ell,P_{zz},\phi) +
\rd\sigma^{\stackrel{\rightarrow}{\Leftarrow}}(e_\ell,P_{zz},\phi)\right]
+\left[\rd\sigma^{\stackrel{\leftarrow}{\Rightarrow}}(e_\ell,P_{zz},\phi) +
\rd\sigma^{\stackrel{\leftarrow}{\Leftarrow}}(e_\ell,P_{zz},\phi)\right]} 
\,.
\label{eq:pmphi}
\end{multline}
Here, the symbol `$\rd\sigma$' denotes a generic differential cross 
section.

For coherent scattering, and to leading order in $\alpha_s$ and in leading 
twist, the expansion in powers of the Bjorken variable $x_{D}$ for the 
deuteron target, and $\tau= t/(4 M_{D}^2)$, where $M_{D}$ is the deuteron 
mass, yields~\cite{theor_deu}
\begin{eqnarray} \nonumber
&&\CalALPM(e_\ell,P_{zz}=+1,\phi) \simeq  -e_\ell
 \frac{x_{D} (2-y)\sqrt{\frac{-t}{Q^2}(1-y)}}{2-2y + y^2} \sin\phi \\ 
\label{eq:App-ALpm}
&\times& \Im\mbox{m} \frac{ G_1 {\cal H}_1 - \frac{1}{3} G_1 {\cal H}_5
- \tau \Big[G_1 {\cal H}_3 + G_3 ({\cal H}_1 - \frac{1}{3}
{\cal H}_5)\Big] + 2\tau^2 G_3 {\cal H}_3}    
{G_1^2 - 2\tau G_1G_3  + 2\tau^2 G_3^2} \\
&\simeq&  -e_\ell
 \frac{x_{D} (2-y)\sqrt{\frac{-t}{Q^2}(1-y)}}{2-2y + y^2}
\frac{\Im\mbox{m}( {\cal H}_1 - \frac{1}{3}{\cal H}_5)}{G_1}\sin\phi\,.
\label{eq:Dpmtilde-short}
\end{eqnarray}
Here, $G_1$ and $G_3$ are deuteron elastic form 
factors~\cite{Deutformfac}. (For comparison with experimental data, the 
actual value of $P_{zz} \ne 1$ must be taken into account in, e.g., 
Eqs.~\ref{eq:App-ALpm}, \ref{eq:Dpmtilde-short}, and those that follow.) 
Equation~\ref{eq:Dpmtilde-short} is obtained neglecting the contributions 
of non-leading terms in $\tau$ in Eq.~\ref{eq:App-ALpm}, which are less 
than 10$\%$ at $-t<0.03\rm\,GeV^2$ (see Fig.~3 in 
Ref.~\cite{deuteron_unpol_draft}). As can be seen from 
Eqs.~\ref{eq:App-ALpm} and~\ref{eq:Dpmtilde-short}, this asymmetry 
involves a different linear combination of the imaginary parts of the 
deuteron CFFs ${\cal H}_1$, ${\cal H}_3$ and ${\cal H}_5$ compared to the 
asymmetry $\CalALUI(\phi)$ (see Eqs. 25-27 in 
Ref.~\cite{deuteron_unpol_draft}). More specifically, any difference 
between these two asymmetries at small values of $-t$ may be ascribed to 
the CFF ${\cal H}_5$. Detailed information about the relations between 
these CFFs and corresponding GPDs can be found in Ref.~\cite{theor_deu}.

Similarly, the beam-charge asymmetry for tensor polarization $P_{zz}$ is 
defined as
\begin{equation}
\CalACPMN(P_{zz},\phi) \equiv \frac
{\Big[ \rd\sigma^{\stackrel{+}{\Rightarrow}}(P_{zz},\phi) + 
\rd\sigma^{\stackrel{+}{\Leftarrow}}(P_{zz},\phi) \Big]
-\Big[ \rd\sigma^{\stackrel{-}{\Rightarrow}}(P_{zz},\phi) + 
\rd\sigma^{\stackrel{-}{\Leftarrow}}(P_{zz},\phi) \Big]}
{\Big[\rd\sigma^{\stackrel{+}{\Rightarrow}}(P_{zz},\phi) + 
\rd\sigma^{\stackrel{+}{\Leftarrow}}(P_{zz},\phi) \Big]
+\Big[\rd\sigma^{\stackrel{-}{\Rightarrow}}(P_{zz},\phi) + 
\rd\sigma^{\stackrel{-}{\Leftarrow}}(P_{zz},\phi) \Big]} \,,
\label{eq:bca}
\end{equation}
where the symbols $+$ ($-$) denote positive (negative) beam charge. For 
coherent scattering, the $\cos\phi$ component in the kinematic expansion 
of Eq.~\ref{eq:bca} is sensitive to the real part of the same linear 
combination of CFFs as that appearing in Eq.~\ref{eq:Dpmtilde-short}:
\begin{equation}
\label{eq:App-ACpm}
\CalACPMN(P_{zz}=+1,\phi) \simeq 
-\frac{x_{D} \sqrt{\frac{-t}{Q^2}(1-y)}}{y} 
\frac{\Re\mbox{e}( {\cal H}_1 - \frac{1}{3}{\cal H}_5)}{G_1} \cos\phi\ \,.
\end{equation}
The different sign of the asymmetry $\CalACPMN(P_{zz}=+1,\phi)$ compared 
to Ref.~\cite{theor_deu} is due to the use of the Trento 
convention~\cite{Trento} in this work, i.e., $\phi = \pi - 
\phi_{{\mbox{\cite{theor_deu}}}}$.

Another single-charge beam-helicity asymmetry,  which  differs from 
$\CalALPM$ ($e_{\ell},P_{zz},\phi$) and $\CalACPMN(e_{\ell},P_{zz},\phi)$, 
involves polarized beam and (longitudinal) tensor polarization of the 
deuteron:
\begin{equation}
\CalALzz(e_\ell,\phi) \equiv \frac
{\rd\sigma^\rightarrow_{\rm zz}(e_\ell,\phi) - 
\rd\sigma^\leftarrow_{\rm zz}(e_\ell,\phi)}
{3\rd\sigma^\rightarrow_{\rmunp}(e_\ell,\phi) + 
3\rd\sigma^\leftarrow_{\rmunp}(e_\ell,\phi)} \,,
\label{equ:tensor}
\end{equation}
with $ \rd\sigma_{\rm zz}$ = $\rd\sigma^\Rightarrow + \rd\sigma^\Leftarrow 
- 2\rd\sigma^0$ and $ \rd\sigma_{\rmunp}$ = 
$\frac{1}{3}(\rd\sigma^\Rightarrow + \rd\sigma^\Leftarrow + 
\rd\sigma^0)$, where $\rd\sigma^0$ represents the cross section for 
deuterons in the $\Lambda = 0$ state. For coherent scattering, the 
asymmetry $\CalALzz(e_\ell,\phi)$ involves a different linear combination 
of the imaginary parts of the deuteron CFFs ${\cal H}_1$, ${\cal H}_3$ and 
${\cal H}_5$ compared to $\CalALPM(e_\ell,P_{zz}=+1,\phi)$ and 
$\CalALUI(\phi)$:
\begin{eqnarray} \nonumber
\CalALzz(e_\ell,\phi) &\simeq& e_\ell
\frac{2 x_{D} (2-y)\sqrt{\frac{-t}{Q^2}(1-y)}}{2-2y + y^2} 
{\sin\phi} \\  \label{eq:App-ALzz}
&\times& {\Im\mbox{m}}
\frac{G_1 {\cal H}_5 +
\tau \left(G_1 {\cal H}_3 + G_3 {\cal H}_1 - \frac{1}{3} G_3 {\cal 
H}_5\right)
-2 \tau^2 G_3 {\cal H}_3 }
 {3 G_1^2 - 4\tau G_1 G_3 + 4\tau^2 G_3^2 }  \\
&\simeq&  e_\ell
\frac{2 x_{D} (2-y)\sqrt{\frac{-t}{Q^2}(1-y)}}{2-2y + y^2}
  \frac{\Im\mbox{m}{\cal H}_5}{3 G_1} \sin\phi \,.
\label{eq:App-ALzz-short}
\end{eqnarray}

Finally, the single-charge asymmetry with respect to longitudinal vector 
polarization of the target is defined as
\begin{multline}
\CalAUL(e_\ell,P_{zz},\phi) \equiv \\
\frac
{\left[\rd\sigma^{\stackrel{\rightarrow}{\Rightarrow}}(e_\ell,P_{zz},\phi) +
\rd\sigma^{\stackrel{\leftarrow}{\Rightarrow}}(e_\ell,P_{zz},\phi)\right]
-\left[\rd\sigma^{\stackrel{\rightarrow}{\Leftarrow}}(e_\ell,P_{zz},\phi) +
\rd\sigma^{\stackrel{\leftarrow}{\Leftarrow}}(e_\ell,P_{zz},\phi)\right]}
{\left[\rd\sigma^{\stackrel{\rightarrow}{\Rightarrow}}(e_\ell,P_{zz},\phi) +
\rd\sigma^{\stackrel{\leftarrow}{\Rightarrow}}(e_\ell,P_{zz},\phi)\right]
+\left[\rd\sigma^{\stackrel{\rightarrow}{\Leftarrow}}(e_\ell,P_{zz},\phi) +
\rd\sigma^{\stackrel{\leftarrow}{\Leftarrow}}(e_\ell,P_{zz},\phi)\right]} 
\,.
\label{eq:ulphi}
\end{multline}
For coherent scattering, in analogy to the previously elaborated 
asymmetries, it reduces to
\begin{eqnarray} \nonumber
& &\CalAUL(e_\ell,P_{zz}=+1,\phi)
\simeq -e_\ell 
\frac{x_{D}\ \sqrt{\frac{-t}{Q^2}(1-y)}\ }{y}
\sin\phi \\  \label{App-AUL}
 &\times& {\Im\mbox{m}}
\frac{\left[ G_1 \widetilde{\cal H}_1 + 
\frac{x_{D}}{2}\ G_2 \left({\cal H}_1 - \frac{1}{3}{\cal H}_5\right) \right] 
- \tau \left( G_3 \widetilde{\cal H}_1 + \frac{x_{D}}{2}\ G_2 
{\cal H}_3 \right)}
{G_1^2 - 2\tau G_1 G_3 + 2\tau^2 G_3^2 }  \\
&\simeq&  -e_\ell
 \frac{x_{D}\ \sqrt{\frac{-t}{Q^2}(1-y)}\ }{y}
  \frac{\Im\mbox{m} \left[ G_1 \widetilde{\cal H}_1 + 
\frac{x_{D}}{2}\ G_2 \left({\cal H}_1 - \frac{1}{3}{\cal H}_5\right) \right]
}{G_1^2} \sin\phi \,.
\label{App-AUL-short}
\end{eqnarray}
Thus, this asymmetry is sensitive to the imaginary part of the Compton 
form factor $\widetilde{\cal H}_1$.

\section{The HERMES experiment}
\label{sec:experiment}
A detailed description of the HERMES spectrometer can be found in 
Ref.~\cite{hermes:spectrometer}. A longitudinally polarized positron or 
electron beam of energy 27.6~GeV was scattered off a longitudinally 
polarized deuterium gas target internal to  the HERA lepton storage ring 
at DESY. The lepton beam was transversely self-polarized by the emission 
of synchrotron radiation~\cite{Sokolov+:1964}. Longitudinal polarization 
of the beam at the target was achieved by a pair of spin rotators in front 
of and behind the experiment~\cite{Buon:1986}. The sign of the beam 
polarization was reversed approximately every two months. Two Compton 
backscattering polarimeters~\cite{TPOL:1994,LPOL:2002} measured 
independently the longitudinal and transverse beam polarizations. The 
average values of the beam polarization for the various running periods 
are given in Table~\ref{tb:table1}; their average fractional systematic 
uncertainty is 2.2$\%$.
\begin{table}[t] \center
\caption{The sign of the beam charge, the luminosity-averaged beam 
polarization and target vector and tensor polarization values, for the 
years 1998-2000 and the integrated luminosity of the data sets used for 
the extraction of the various asymmetry amplitudes (see 
Table~\ref{tb:table_asymm}) on a longitudinally polarized deuterium 
target. The uncertainties for the polarizations are given in the 
text.}
\begin{tabular}{ccrrccccr}
\noalign{\smallskip}
\hline
    & Lepton &  \multicolumn{2}{c}{Beam} & 
\multicolumn{2}{c}{\hspace{0.4cm} Target} 
&\multicolumn{3}{c}{\hspace{0.4cm}Luminosity [pb$^{-1}$]} \\
Year&Charge& \multicolumn{2}{c}{\hspace{0.1cm} Polarization} & 
\multicolumn{2}{c}{\hspace{0.35cm} Polarization} & \hspace{0.25cm}
$\CalALPM$ & \hspace{0.25cm}$\CalACPM$ & \hspace{0.25cm} $\CalALzz$ \\
& & & & \hspace{0.25cm} $P_{z}$ & $P_{zz}$ & \hspace{0.5cm}($\CalAUL$, 
$\CalALL$) & \hspace{0.5cm}($\CalAP$, $\CalACP$) & \\
\noalign{\smallskip}
\hline
1998&$e^-$ & $- \, 0.509$ & & \hspace{0.25cm} $\pm$0.856 & 
\hspace{0.25cm} +\,0.827 & & \hspace{0.25cm} 26.2 &\\
1999&$e^+$ & $- \, 0.547,$ & +\,0.518 & \hspace{0.25cm} $\pm$0.832 & 
\hspace{0.25cm} +\,0.827 & \hspace{0.25cm} 29.7 & \hspace{0.25cm} 14.2 & 
29.7\\
2000&$e^+$ & $- \, 0.537,$ &+\,0.524 & \hspace{0.25cm} $- \,0.840, + \, 
0.851$ & \hspace{0.25cm} +\,0.827 & \hspace{0.25cm} 125.8 & 
\hspace{0.25cm} 43.5 & 125.8\\
2000&$e^+$ & $- \, 0.542,$ &+\,0.525 & $- \, 0.010$ & \hspace{0.25cm} 
$-\, 1.656$ & & & 22.7\\
\hline
Sum& & & & & & \hspace{0.25cm} $155.5$ & \hspace{0.25cm} $83.9$ & 
$178.2$\\
\hline
\end{tabular}
\label{tb:table1}
\end{table}

The target cell was filled with nuclear-polarized atoms from an atomic 
beam source based on Stern--Gerlach separation with radio-frequency 
hyperfine transitions~\cite{hermes:ABS}. The polarization and atomic 
fraction of the target gas were continuously 
monitored~\cite{hermes:BRP,hermes:TGA}. Most of the longitudinally 
polarized deuterium data were recorded with average vector polarizations 
$0.851\pm0.031$ and $-0.840\pm0.028$, and with an average tensor 
polarization of $0.827\pm0.027$~\cite{hermes:target} (corresponding to a 
small population of the $\Lambda$ = 0 state). The extraction of 
$\CalALzz(\phi)$ employed the fraction of the data taken in the year 2000 
recorded with a tensor-polarized deuterium target where deuterons in the 
$\Lambda = 0$ state were injected into the target cell, resulting in an 
average tensor polarization of $-1.656\pm0.049$ with negligible vector 
polarization ($-0.010\pm0.026$). The amount of data accumulated for each 
lepton beam charge and sign of the polarization are summarized in 
Table~\ref{tb:table_lumi}.
\begin{table} \center
\caption{The integrated luminosity of the data used for the extraction of 
various asymmetry amplitudes (see Table~\ref{tb:table_asymm}) on a 
longitudinally 
polarized deuterium target for each lepton beam charge and sign of the 
polarization.}
\begin{tabular}{ccc}
\noalign{\smallskip}
\hline
Lepton Charge & \hspace{0.5cm} Sign of the Beam Polarization & 
\hspace{0.5cm} Luminosity 
[pb$^{-1}$] \\
\noalign{\smallskip}
\hline
$e^-$ & negative & 26.2 \\
$e^-$ & positive & \\
$e^+$ & negative & 75.4 \\
$e^+$ & positive & 102.8 \\
\hline
\end{tabular}
\label{tb:table_lumi}
\end{table}

The scattered leptons and produced particles were detected by the HERMES 
spectrometer in the polar angle range $0.04$~rad~$< \theta < 0.22$~rad. 
The average lepton identification efficiency was at least 98$\%$ with 
hadron contamination of less than 1$\%$.

\section{Event selection}
\label{sec:event_yield}
The data sets used in the extraction of the various asymmetries reported 
here are given in Table~\ref{tb:table1}. In this analysis, it was required 
that events contain exactly one charged-particle track identified as a 
lepton with the same charge as the beam lepton, and one photon producing 
an energy deposition $E_{\gamma} > 5\rm\,GeV$ ($> 1\rm\,MeV$) in the 
calorimeter (preshower detector). The following kinematic requirements 
were imposed on the events, as calculated from the four-momenta of the 
incoming and outgoing lepton: $1\rm\,GeV^2$ $< Q^2 <$ $10\rm\,GeV^2 $, 
$W_N^2 >9\rm\,GeV^2$, $\nu<22\rm\,GeV$ and $ 0.03 < x_N <0.35 $, where 
$W_N^2 = M_N^2 + 2M_N\nu - Q^2$ and $x_N = Q^2/(2M_{N} \nu)$. For the 
nucleonic mass $M_{N}$, the proton mass was used in all kinematic 
constraints on event selection, even at small values of $-t$ where 
coherent reactions on the deuteron are dominant, because the experiment 
does not distinguish between coherent and incoherent scattering and the 
latter dominates over most of the kinematic range. Monte Carlo studies 
have shown that this choice has little effect on the result~\cite{Bernie}. 
In order to reduce background from the decay of neutral mesons, the angle 
between the laboratory three-momenta of the real and virtual photons was 
limited to $\theta_{\gamma^* \gamma} < 45$\,mrad. The minimum angle 
requirement $\theta_{\gamma^* \gamma} > 5$\,mrad was chosen according to 
Monte Carlo simulations in order to ensure that the azimuthal angle $\phi$ 
remains well-defined while accounting for the finite angular resolution of 
the spectrometer.

An `exclusive' event sample was selected by requiring the squared missing 
mass $M_X^2$ to be close to the squared nucleon mass $M_N^2$, where 
$M_X^2$ is defined as $M_X^2 = (q + P_N - q^\prime)^2$ with $P_N = 
(M_N,0,0,0)$ and $q^\prime$ the four-momentum of the real photon. The 
exclusive region is defined as $-(1.5)^2{\rm\,GeV}^2 < M_X^2 
<(1.7)^2{\rm\,GeV}^2$ to minimize background from deep-inelastic 
scattering fragmentation processes, while maintaining reasonable 
efficiency~\cite{Frank}. 

As the recoiling target nucleon or nucleus was undetected, the Mandelstam 
variable $t$ was reconstructed from the measured four-momenta of the 
scattered lepton and the detected photon. The resolution in the photon 
energy from the calorimeter is inadequate for a precise determination of 
$t$. Hence for events selected in the exclusive region in $M_X^2$, the 
reaction is assumed to take place on a nucleon and the final state is 
assumed to contain only the scattered lepton, the real photon and the 
nucleon that was left intact ($e \, N \to e \, N \, \gamma$). This 
allows $t$ to be calculated with improved resolution using only the 
photon direction and the lepton four-momentum~\cite{hermes_bca_2006}:
\begin{equation}
t = \frac{-Q^2 - 2 \, \nu \, (\nu - \sqrt{\nu^2 + Q^2} \, 
\cos\theta_{\gamma^* \gamma })}
{1 + \frac{1}{M_N} \, (\nu - \sqrt{\nu^2 + Q^2} \, \cos\theta_{\gamma^* 
\gamma})} \, .
\label{tc}
\end{equation}
The error caused by applying this expression to incoherent events with a 
nucleon excited to a resonance in the final state is accounted for in the 
Monte Carlo simulation that is used to calculate the fractional 
contribution of background processes per kinematic bin. This simulation 
also demonstrated that this method is applicable also to coherent events. 
A further restriction, $-t < 0.7\rm\,GeV^2$, is used in the selection of 
exclusive events in order to reduce background.

The exclusive sample comprises coherent and incoherent scattering, 
including resonance excitation. Over most of the kinematic range 
incoherent scattering dominates. The events from coherent scattering off 
the deuteron are concentrated at small values of $-t$. The Monte Carlo 
simulation showed that requiring $-t < 0.06\rm\,GeV^2$ enhances the 
relative contribution of the coherent process from 20$\%$ to 40$\%$ in the 
data sample. Requiring $-t < 0.01\rm\,GeV^2$ can further enhance the 
coherent contribution to 66$\%$, but only at the cost of a rapidly 
decreasing yield. The first bin defined in Section~\ref{sec:results_pol1} 
covering the range $-t < 0.06\rm\,GeV^2$ is sensitive to coherent effects.

\section{Extraction formalism}
\label{sec:formalism}
The simultaneous extraction of Fourier amplitudes of beam-charge and 
beam-helicity asymmetries combining data collected during various running 
periods at HERMES for both beam charges and helicities on unpolarized 
hydrogen or deuterium targets is described in 
Refs.~\cite{deuteron_unpol_draft,proton_unpol_draft}. It is based on the 
maximum likelihood technique~\cite{MML}, which provides a bin-free fit in 
the azimuthal angle $\phi$ (see Ref.~\cite{hermes_ttsa} for details). In 
this paper, data taken with a longitudinally polarized deuterium target 
were analyzed with a similar technique. In the fit, event weights were 
introduced to account for luminosity imbalances with respect to beam 
charge and polarization.

Because the target polarization was longitudinal with respect to the 
direction of the incoming beam, the data also contain contributions 
arising from the small transverse polarization with respect to the 
direction of the virtual photon. This $6\% - 12\%$ transverse component of 
the target polarization, depending on the kinematic conditions of each 
bin, was neglected in the formalism presented. Hence, the extracted 
Fourier components contain contributions from this transverse component. 
However, mainly non-leading (higher-twist) amplitudes are affected by this 
choice. These effects are estimated from the measurement of the 
transverse-target-spin asymmetries at HERMES~\cite{hermes_ttsa} to be less 
than 0.008 on a proton target, and hence are expected to be negligible 
compared with the uncertainties here.

\subsection{Single-charge formalism}
\label{single_charge_formalism}
Data collected with an e$^-$ beam and a polarized deuterium target were 
not used for the extraction of harmonics of $\CalALPM$, $\CalAUL$ and 
$\CalALL$ because only negative beam polarization is available for this 
charge. Hence, Fourier amplitudes of the three {\it single-charge} 
asymmetries $\CalALPM(e_\ell=+1,P_{zz}, \phi)$, 
$\CalAUL(e_\ell=+1,P_{zz},\phi)$ and $\CalALL(e_\ell=+1,P_{zz},\phi)$, 
defined respectively in Eqs.~\ref{eq:pmphi}, \ref{eq:ulphi}, and 
\ref{eq:sigma00_assall_beg}, are simultaneously extracted using data from 
scattering of a longitudinally polarized positron beam off a 
longitudinally polarized deuterium target. 

The distribution in the expectation value of the yield can be written as
\begin{multline}
\rd\langle\intN\rangle(e_\ell=+1,P_\ell,P_{z},P_{zz},\phi)=
\Lumi(e_\ell=+1,P_\ell,P_{z},P_{zz})\,\eta(\phi) \\
\times \rd\CULPM(e_\ell=+1,P_{zz},\phi) 
\Big[1+P_\ell\CalALPM(e_\ell=+1,P_{zz},\phi) \\
+ P_z\CalAUL(e_\ell=+1,P_{zz},\phi)+P_\ell P_z\CalALL
(e_\ell=+1,P_{zz},\phi)\Big]\, ,
\label{eq:A2}
\end{multline}
where $\Lumi$ denotes the integrated luminosity and $\eta$ the detection 
efficiency. The cross section for the production of real photons by {\em 
unpolarized} positrons on a tensor-polarized deuterium target with 
vanishing vector polarization is given by
\begin{align}
\label{eq:sigmaulpm}
& \begin{aligned}[b] \, \rd\CULPM(e_\ell=+1,P_{zz},\phi) &\equiv 
\frac{1}{4}\Big[\rd\sigma^{\stackrel{\rightarrow}{\Rightarrow}+}
(P_{zz},\phi) +
\rd\sigma^{\stackrel{\leftarrow}{\Leftarrow}+}(P_{zz},\phi) \\
& + \rd\sigma^{\stackrel{\leftarrow}{\Rightarrow}+}(P_{zz},\phi) 
+ \rd\sigma^{\stackrel{\rightarrow}{\Leftarrow}+}(P_{zz},\phi) 
\Big] \end{aligned} \\
& = K \Bigg\{\frac{K_{\rm BH}}{{\cal P}_{1}(\phi){\cal P}_{2}(\phi)}
\Big[\sum_{n=0}^2c_{n,\rm unp}^{\rm BH}\cos(n\phi)+\frac{1}{2}P_{zz}
\sum_{n=0}^2(c_{n,\rm unp}^{\rm BH}-c_{n,\rm LLP}^{\rm BH})
\cos(n\phi)\Big] \nonumber \\
& +K_{\rm DVCS}\Big[\sum_{n=0}^2c_{n,\rm unp}^{\rm DVCS}\cos(n\phi)+
\frac{1}{2}P_{zz}\sum_{n=0}^2(c_{n,\rm unp}^{\rm DVCS}-
c_{n,\rm LLP}^{\rm DVCS})\cos(n\phi)\Big] \nonumber \\
& -\frac{K_{\rm I}}{{\cal P}_{1}(\phi){\cal P}_{2}(\phi)}
\Big[\sum_{n=0}^3c_{n,\rm unp}^{\rm I}\cos(n\phi)+\frac{1}{2}P_{zz}
\sum_{n=0}^3(c_{n,\rm unp}^{\rm I}-c_{n,\rm LLP}^{\rm I})
\cos(n\phi)\Big]\Bigg\}\, , 
\end{align}
where $K = \frac{x_{D} \, e^6}{32 \, (2 \pi)^4 \, 
Q^4\sqrt{1+\varepsilon^2}}$ is a common kinematic factor.

The single-charge asymmetries appearing in Eq.~\ref{eq:A2} are expanded in 
terms of the same Fourier harmonics used in the expansion of the cross 
section in Eqs.~\ref{eq:moments-BHVT}--\ref{eq:moments-IVT} and in the 
numerators appearing in Eqs.~\ref{eq:sigma00_assalpm_mid}, 
\ref{eq:sigma00_assaul_mid}, and \ref{eq:sigma00_assall_mid}:
\begin{align} \nonumber
& \CalALPM(e_\ell=+1,P_{zz},\phi) = 
\frac{K}{\rd\CULPM(e_\ell=+1,P_{zz},\phi)} \\
& \times \Bigg\{K_{\rm DVCS}\Big[s_{1,\rm unp}^{\rm DVCS}\sin \phi+
\frac{1}{2}P_{zz}\, (s_{1,\rm unp}^{\rm DVCS}-s_{1,\rm LLP}^{\rm DVCS})
\sin \phi \Big] \nonumber \\ \label{eq:sigma00_assalpm_mid}
& -\frac{K_{\rm I}}{{\cal P}_{1}(\phi){\cal P}_{2}(\phi)}
\Big[\sum_{n=1}^2s_{n,\rm unp}^{\rm I}\sin(n\phi)+\frac{1}{2}P_{zz}
\sum_{n=1}^2(s_{n,\rm unp}^{\rm I}-
s_{n,\rm LLP}^{\rm I})\sin(n\phi)\Big]\Bigg\} \\
&\hspace{3.2cm} \simeq \sum_{n=1}^2 
\ALPM^{\sin(n\phi)}(e_\ell=+1,P_{zz})\sin(n\phi) \, ,
\label{eq:sigma00_assalpm}
\end{align}
\begin{eqnarray} \nonumber
& & \CalAUL(e_\ell=+1,P_{zz},\phi) = 
\frac{K}{\rd\CULPM(e_\ell=+1,P_{zz},\phi)} \\
& & \times {}\Bigg\{K_{\rm DVCS}
\sum_{n=1}^2s_{n,\rm LP}^{\rm DVCS}\sin(n\phi)
-\frac{K_{\rm I}}{{\cal P}_{1}(\phi){\cal P}_{2}(\phi)}
\sum_{n=1}^3s_{n,\rm LP}^{\rm I}\sin(n\phi)\Bigg\}{} 
\label{eq:sigma00_assaul_mid} \\
& & \hspace{3.2cm} \simeq \sum_{n=1}^3 
\AUL^{\sin(n\phi)}(e_\ell=+1,P_{zz})\sin(n\phi) \, ,
\label{eq:sigma00_assaul}
\end{eqnarray}
\begin{eqnarray}
\label{eq:sigma00_assall_beg} 
& & \CalALL(e_\ell=+1,P_{zz},\phi) \equiv 
\frac{1}{4\,\rd\CULPM(e_\ell=+1,P_{zz},\phi)} \nonumber \\
& & \times \Bigg\{\Big[\rd\sigma^{\stackrel{\rightarrow}{\Rightarrow}+}
(P_{zz},\phi) + 
\rd\sigma^{\stackrel{\leftarrow}{\Leftarrow}+}(P_{zz},\phi)\Big] 
- \Big[\rd\sigma^{\stackrel{\leftarrow}{\Rightarrow}+} (P_{zz},\phi) +
\rd\sigma^{\stackrel{\rightarrow}{\Leftarrow}+}(P_{zz},\phi)\Big] \Bigg\} \\
& & \hspace{3.2cm} = \frac{K}{\rd\CULPM(e_\ell=+1,P_{zz},\phi)} \nonumber \\
& & \times \Bigg\{\frac{K_{\rm BH}}{{\cal P}_{1}(\phi){\cal P}_{2}(\phi)}
\sum_{n=0}^1c_{n,\rm LP}^{\rm BH}\cos(n\phi)+K_{\rm DVCS}
\sum_{n=0}^1c_{n,\rm LP}^{\rm DVCS}\cos(n\phi) \nonumber \\
& & -\frac{K_{\rm I}}{{\cal P}_{1}(\phi){\cal P}_{2}(\phi)}
\sum_{n=0}^2c_{n,\rm LP}^{\rm I}\cos(n\phi)\Bigg\} 
\label{eq:sigma00_assall_mid} \\
& & \hspace{3.2cm}\simeq \sum_{n=0}^2 
\ALL^{\cos(n\phi)}(e_\ell=+1,P_{zz})\cos(n\phi) \, .
\label{eq:sigma00_assall}
\end{eqnarray}
The approximation in Eqs.~\ref{eq:sigma00_assalpm}, 
\ref{eq:sigma00_assaul}, and \ref{eq:sigma00_assall} is due to the 
truncation of terms in the Fourier series arising from the azimuthal 
dependences in the common denominator and the lepton propagators of 
Eqs.~\ref{eq:sigma00_assalpm_mid}, \ref{eq:sigma00_assaul_mid}, and 
\ref{eq:sigma00_assall_mid}. The Fourier coefficients of the expansion of 
the asymmetries are hereafter called asymmetry amplitudes. Although these 
asymmetry amplitudes differ from the coefficients appearing in 
Eqs.~\ref{eq:moments-BHVT}--\ref{eq:moments-IVT} and 
Eqs.~\ref{eq:sigma00_assalpm_mid}, \ref{eq:sigma00_assaul_mid}, and 
\ref{eq:sigma00_assall_mid}, they may provide similar information in the 
comparison of model predictions with data.

\subsection{Single-beam-helicity formalism}
\label{single_beam_helicity}
In order to extract more information on various combinations of 
Fourier coefficients in Eqs.~\ref{eq:moments-BHVT}--\ref{eq:moments-IVT}, 
it is possible to use data collected with negative polarization of the 
e$^-$ beam in conjunction with the subset of positron data with the same 
sign of the beam polarization. In this case, another set of Fourier 
coefficients of the {\it single-beam-helicity} asymmetries 
$\CalACPM(P_\ell,P_{zz},\phi)$, $\CalAP(P_\ell,P_{zz},\phi)$ and 
$\CalACP(P_\ell,P_{zz},\phi)$ can be simultaneously extracted, where the 
subscript 
$\mathrm{{\stackrel{C}{\leftarrow}}{\stackrel{\Leftarrow}{\Rightarrow}}}$ 
indicates the charge asymmetry for a lepton beam with negative 
polarization on a longitudinally polarized deuterium target with 
vanishing net vector polarization. The subscript 
$\mathrm{{\stackrel{0}{\leftarrow}}L}$ indicates the asymmetry with 
respect to longitudinal vector target polarization for a charge-averaged 
lepton beam again with negative beam polarization. Similarly, the 
subscript $\mathrm{{\stackrel{C}{\leftarrow}}L}$ indicates the double 
asymmetry with respect to lepton charge and longitudinal vector target 
polarization.

The azimuthal distribution in the expectation value of the yield in this 
case can be written as
\begin{multline}
\rd\langle\intN\rangle(e_\ell,P_\ell,P_{z},P_{zz},\phi)=\Lumi(e_\ell,P_\ell,P_{z},P_{zz})
\,\eta(\phi)\,\rd\CLCPM(P_\ell,P_{zz},\phi)\\
\times 
\Big[1+e_\ell\CalACPM(P_\ell,P_{zz},\phi)+P_z\CalAP(P_\ell,P_{zz},\phi)+
e_\ell P_z\CalACP(P_\ell,P_{zz},\phi)\Big]\, .\label{eq:A3}
\end{multline}
Here, the cross section $\rd\CLCPM(P_\ell,P_{zz},\phi)$ for production of 
real photons by a charge-averaged polarized lepton beam on a 
tensor-polarized deuterium target with vanishing vector polarization is 
defined as
\begin{align}
& \begin{aligned}[b] \, \rd\CLCPM(P_\ell,P_{zz},\phi) & \equiv 
\frac{1}{4} \Big[\rd\sigma^{{\stackrel{\leftarrow}{\Rightarrow}}+}
(P_\ell,P_{zz},\phi) 
+ \rd\sigma^{{\stackrel{\leftarrow}{\Leftarrow}}+}
(P_\ell,P_{zz},\phi) \\
& + \rd\sigma^{{\stackrel{\leftarrow}{\Rightarrow}}-}
(P_\ell,P_{zz},\phi) + 
\rd\sigma^{{\stackrel{\leftarrow}{\Leftarrow}}-}
(P_\ell,P_{zz},\phi) \Big] \end{aligned} \\
& = K \Bigg\{ \frac{K_{\rm BH}}{{\cal P}_{1}(\phi){\cal 
P}_{2}(\phi)}\bigg[\sum_{n=0}^2 c_{n,\rm unp}^{\rm BH}
\cos(n\phi)+\frac{1}{2}P_{zz}\sum_{n=0}^2
(c_{n,\rm unp}^{\rm BH}-c_{n,\rm LLP}^{\rm BH})\cos(n\phi)\bigg]{} 
\nonumber 
\label{eq:sigmaclcpm} \\
& + K_{\rm DVCS}\bigg[\sum_{n=0}^2 c_{n,\rm unp}^{\rm DVCS} 
\cos(n\phi)+P_\ell
\sum_{n=1}^2 s_{n,\rm unp}^{\rm DVCS} \sin(n\phi){} \nonumber \\
& + \frac{1}{2}P_{zz}\bigg(\sum_{n=0}^2 
(c_{n,\rm unp}^{\rm DVCS}-c_{n,\rm LLP}^{\rm DVCS})
\cos(n\phi)+P_\ell\sum_{n=1}^2 (s_{n,\rm unp}^{\rm DVCS}-
s_{n,\rm LLP}^{\rm DVCS})\sin(n\phi)\bigg)\bigg]\Bigg\}\,. \end{align}
Then the single-beam-helicity asymmetries appearing in Eq.~\ref{eq:A3} are 
expressed as
\begin{align}
& \begin{aligned}[b] \CalACPM(P_\ell,P_{zz},\phi) & \equiv \frac{1}{4 \, 
\rd\CLCPM(P_\ell,P_{zz},\phi)} \\
& \times \Bigg\{\Big[\rd\sigma^{{\stackrel{\leftarrow}{\Rightarrow}}+}
(P_\ell,P_{zz},\phi)
+ \rd\sigma^{{\stackrel{\leftarrow}{\Leftarrow}}+}
(P_\ell,-P_{z},P_{zz},\phi) \Big] \\
& - \Big[\rd\sigma^{{\stackrel{\leftarrow}{\Rightarrow}}-}
(P_\ell,P_{zz},\phi)
+ \rd\sigma^{{\stackrel{\leftarrow}{\Leftarrow}}-}
(P_\ell,P_{zz},\phi) \Big] \Bigg\} \label{eq:acpm_beg} \end{aligned} \\
& \hspace{2.5cm} = \frac{K}{\rd\CLCPM(P_\ell,P_{zz},\phi)} \label{eq:acpm} 
\nonumber \\
& \times \Bigg\{-\frac{K_{\rm I}}{{\cal P}_{1}(\phi){\cal P}_{2}(\phi)}
\bigg[\sum_{n=0}^3 c_{n,\rm unp}^{\rm I}\cos(n\phi)+P_\ell 
\sum_{n=1}^2 s_{n,\rm unp}^{\rm I}\sin(n\phi) \nonumber \\
& + \frac{1}{2}P_{zz}\bigg(\sum_{n=0}^3(c_{n,\rm unp}^{\rm I}-
c_{n,\rm LLP}^{\rm I})\cos(n\phi)+P_\ell\sum_{n=1}^2(s_{n,\rm unp}^{\rm I}-
s_{n,\rm LLP}^{\rm I})\sin(n\phi)\bigg)\bigg]\Bigg\} \\
& \hspace{2.5cm} \simeq \sum_{n=0}^3 \ACPMM^{\cos(n\phi)}(P_{zz})\cos(n\phi) 
+ P_\ell \sum_{n=1}^2 \ACPMM^{\sin(n\phi)}(P_{zz})\sin(n\phi)\, , 
\label{eq:acpm_end}
\end{align}
\begin{align}
& \begin{aligned}[b] \CalAP(P_\ell,P_{zz},\phi) & \equiv \frac{1}{4 \, 
\rd\CLCPM(P_\ell,P_{zz},\phi)} \\
& \times \Bigg\{\Big[\rd\sigma^{{\stackrel{\leftarrow}{\Rightarrow}}+}
(P_\ell,P_{zz},\phi)
+ \rd\sigma^{{\stackrel{\leftarrow}{\Rightarrow}}-}
(P_\ell,P_{zz},\phi) \Big] \\
& - \Big[\rd\sigma^{{\stackrel{\leftarrow}{\Leftarrow}}+}
(P_\ell,P_{zz},\phi)
+ \rd\sigma^{{\stackrel{\leftarrow}{\Leftarrow}}-}
(P_\ell,P_{zz},\phi) \Big] \Bigg\} \label{eq:ap_beg} \end{aligned} \\
& \hspace{2.5cm} =  \frac{K}{\rd\CLCPM(P_\ell,P_{zz},\phi)} \label{eq:ap} 
\nonumber \\
& \times \Bigg\{
\frac{K_{\rm BH}}{{\cal P}_{1}(\phi){\cal P}_{2}(\phi)}\Big[
P_\ell\sum_{n=0}^1 c_{n,\rm LP}^{\rm BH}\cos(n\phi)\Big] \nonumber \\
& + K_{\rm DVCS}\Big[P_\ell\sum_{n=0}^1 
c_{n,\rm LP}^{\rm DVCS}\cos(n\phi)+
\sum_{n=1}^2s_{n,\rm LP}^{\rm DVCS}\sin(n\phi)\Big]\Bigg\} \\
& \hspace{2.5cm} \simeq P_\ell \sum_{n=0}^1 \AP^{\cos(n\phi)}(P_{zz})\cos(n\phi)
+ \sum_{n=1}^2 \AP^{\sin(n\phi)}(P_{zz})\sin(n\phi) \, ,
\label{eq:ap_end}
\end{align}
\begin{align}
& \begin{aligned}[b] \CalACP(P_\ell,P_{zz},\phi) & \equiv \frac{1}{4 \, 
\rd\CLCPM(P_\ell,P_{zz},\phi)} \\
& \times \Bigg\{\Big[ \rd\sigma^{{\stackrel{\leftarrow}{\Rightarrow}}+}
(P_\ell,P_{zz},\phi)
+ \rd\sigma^{{\stackrel{\leftarrow}{\Leftarrow}}-}
(P_\ell,P_{zz},\phi) \Big] \\
& - \Big[ \rd\sigma^{{\stackrel{\leftarrow}{\Leftarrow}}+}
(P_\ell,P_{zz},\phi)
+ \rd\sigma^{{\stackrel{\leftarrow}{\Rightarrow}}-}
(P_\ell,P_{zz},\phi) \Big] \Bigg\} \label{eq:acp_beg} \end{aligned} \\
& \hspace{2.5cm} = \frac{K}{\rd\CLCPM(P_\ell,P_{zz},\phi)} \label{eq:acp} 
\nonumber \\
& \times \Bigg\{-\frac{K_{\rm I}}{{\cal P}_{1}(\phi){\cal P}_{2}(\phi)} 
\Big[P_\ell\sum_{n=0}^2c_{n,\rm LP}^{\rm I}\cos(n\phi)+
\sum_{n=1}^3s_{n,\rm LP}^{\rm I}\sin(n\phi)\Big]\Bigg\} \\
& \hspace{2.5cm} \simeq  P_\ell \sum_{n=0}^2 \ACP^{\cos(n\phi)}(P_{zz})\cos(n\phi)
+ \sum_{n=1}^3 \ACP^{\sin(n\phi)}(P_{zz})\sin(n\phi) \, .
\label{eq:acp_end}
\end{align}
All the asymmetries defined in this paper are summarized in 
Table~\ref{tb:table_asymm}.

\begin{table}[h!]
\caption{
Extracted beam-helicity, beam-charge and target-spin asymmetries on a 
polarized deuterium target. The symbol $\blacksquare$ marks which data 
taken under certain experimental conditions (beam polarization, beam 
charge and target polarization state) are available for the construction 
of the respective asymmetry. The $-$ or $+$ indicates the sign with which 
the corresponding yield enters the numerator of the asymmetry. For the 
case that the target is populated with deuterons in the state $\Lambda=\pm 
1$, the ideal target polarizations are $P_z=\pm 1$ and $P_{zz}=1$, while 
for the case $\Lambda=0$, $P_z=0$ and $P_{zz}=-2$. The sensitivity of 
coherent scattering to the corresponding Compton form factors or BH 
amplitude is indicated.}

\centering{
\begin{tabular}{l|c|ccc|ccccc|ccc|c}
\noalign{\smallskip}
& & \multicolumn{3}{c|}{{\small Lepton charge}} & \multicolumn{5}{c|}{{\small Target population (deuterons)}} & \multicolumn{3}{c|}{{\small Beam helicity}}& \\\hline
& & & &   & {\small $\Lambda=+1$} & & {\small $\Lambda=-1$} & & {\small $\Lambda=0$} & {\small $\lambda=+1$} & & {\small $\lambda=-1$} &Coherent\\
& & $\;\;+1\;\;$ & & $\;\;-1\;\;$ & $\Rightarrow$& & $\Leftarrow$ & &    0        & $\rightarrow$& & $\leftarrow$ & sensitivity\\ \hline
\multirow{4}{*}{\rotatebox{90}{\mbox{\small Single-charge}}}
& $\CalALPM$ & $\blacksquare$ & & & $\blacksquare$ & $+$ & $\blacksquare$ & & &  $\blacksquare$ & $-$ & $\blacksquare$ & $\Im\mathrm{m}(\mathcal{H}_1,\mathcal{H}_5)$\\
& $\CalAUL$ & $\blacksquare$ &  & & $\blacksquare$ & $-$ & $\blacksquare$ & & &  $\blacksquare$ & $+$ & $\blacksquare$ & $\Im\mathrm{m}(\mathcal{\widetilde{H}}_1)$\\
& $\CalALL$ & $\blacksquare$ &  &  & $\blacksquare$ & $-$ & $\blacksquare$ & & &  $\blacksquare$ & $-$ & $\blacksquare$ & (BH)\\
& $\CalALzz$ & $\blacksquare$ &  &  & $\blacksquare$ & $+$ & $\blacksquare$ & $-$ & $\blacksquare$ &  $\blacksquare$ & $-$ & $\blacksquare$ &$\Im\mathrm{m}(\mathcal{H}_5)$\\\hline
\multirow{3}{*}{\rotatebox{90}{\mbox{\small Single-helicity}}}
& $\CalACPM$ & $\blacksquare$ & $-$ & $\blacksquare$ & $\blacksquare$ & $+$ & $\blacksquare$ & & & &  & $\blacksquare$ & $\Im\mathrm{m}/\Re\mathrm{e}(\mathcal{H}_1,\mathcal{H}_5)$\\
& $\CalAP$ & $\blacksquare$ & $+$ & $\blacksquare$ & $\blacksquare$ & $-$ & $\blacksquare$ & & & & & $\blacksquare$ & (BH) \\
& $\CalACP$ & $\blacksquare$ & $-$ & $\blacksquare$ & $\blacksquare$ & $-$ & $\blacksquare$ & & &  & & $\blacksquare$ &$\Im\mathrm{m}/\Re\mathrm{e}(\mathcal{\widetilde{H}}_1)$\\
\end{tabular}
}
\label{tb:table_asymm}
\end{table}

\section{Background corrections and systematic uncertainties}
\label{sec:background}
The asymmetry amplitudes are corrected for background contributions, 
mainly decays to two photons of semi-inclusive neutral mesons, using the 
method described in detail in Ref.~\cite{hermes_ttsa}. The average 
contribution from semi-inclusive background is 4.6$\%$. The contribution 
of exclusive pions is neglected, as it is found to be less than 0.7$\%$ in 
each kinematic bin, supported by studies of HERMES data~\cite{Arne}. After 
applying this correction, the resulting asymmetry amplitudes are expected 
to originate from coherent and incoherent photon production, the latter 
possibly including nucleon excitation.

The dominant contributions to the total systematic uncertainty are the 
effects of the limited spectrometer acceptance and from the finite bin 
widths used for the final presentation of the results. The latter 
originates from the difference of the amplitudes integrated over one bin 
in all kinematic variables, compared to the asymmetry amplitudes 
calculated at the average values of the kinematic variables. The combined 
contribution to the systematic uncertainty from limited spectrometer 
acceptance, finite bin width, and the alignment of the spectrometer 
elements with respect to the beam is determined from a Monte Carlo  
simulation using a convenient parameterization~\cite{GPD:PROTON} of the 
VGG model~\cite{Vanderhaeghen:1999xj} (see details in 
Ref.~\cite{deuteron_unpol_draft}). Five GPD model variants are considered, 
including only incoherent processes on the proton and neutron. In each 
kinematic bin, the resulting systematic uncertainty is defined as the 
root-mean-square average of the five differences between the asymmetry 
amplitude extracted from the Monte Carlo data and the corresponding model 
predictions calculated analytically at the mean kinematic values of that 
bin. In the case of the single-charge beam-helicity asymmetry, all five 
models overpredict the magnitudes of the $\sin \phi$ harmonics by about a 
factor of two, leading to a probable overestimate of this contribution to 
the uncertainties. The other source of uncertainty is associated with the 
background correction. For asymmetries involving target vector 
polarization, no systematic uncertainty due to luminosity is assigned. 
This is legitimate because the luminosity does not depend on the target 
polarization, the target polarization flips rapidly compared to changes in 
luminosity, and beam polarization dependent weights are assigned to each 
event in the extraction. There is an additional overall scale uncertainty 
arising from the uncertainty in the measurement of the beam and/or target 
polarizations. Not included is any contribution due to additional QED 
vertices, as for the case of polarized target and polarized beam the most 
significant of these has been estimated to be negligible~\cite{Afanasev}. 
The total systematic uncertainty in a kinematic bin is determined by 
adding quadratically all contributions to the systematic uncertainty for 
that bin.

\section{Results}
\label{sec:results_pol1}
\subsection{Single- and double-spin asymmetries}
\label{single_double_asymm}
\begin{figure}[!tb]
\includegraphics[width=1\columnwidth]{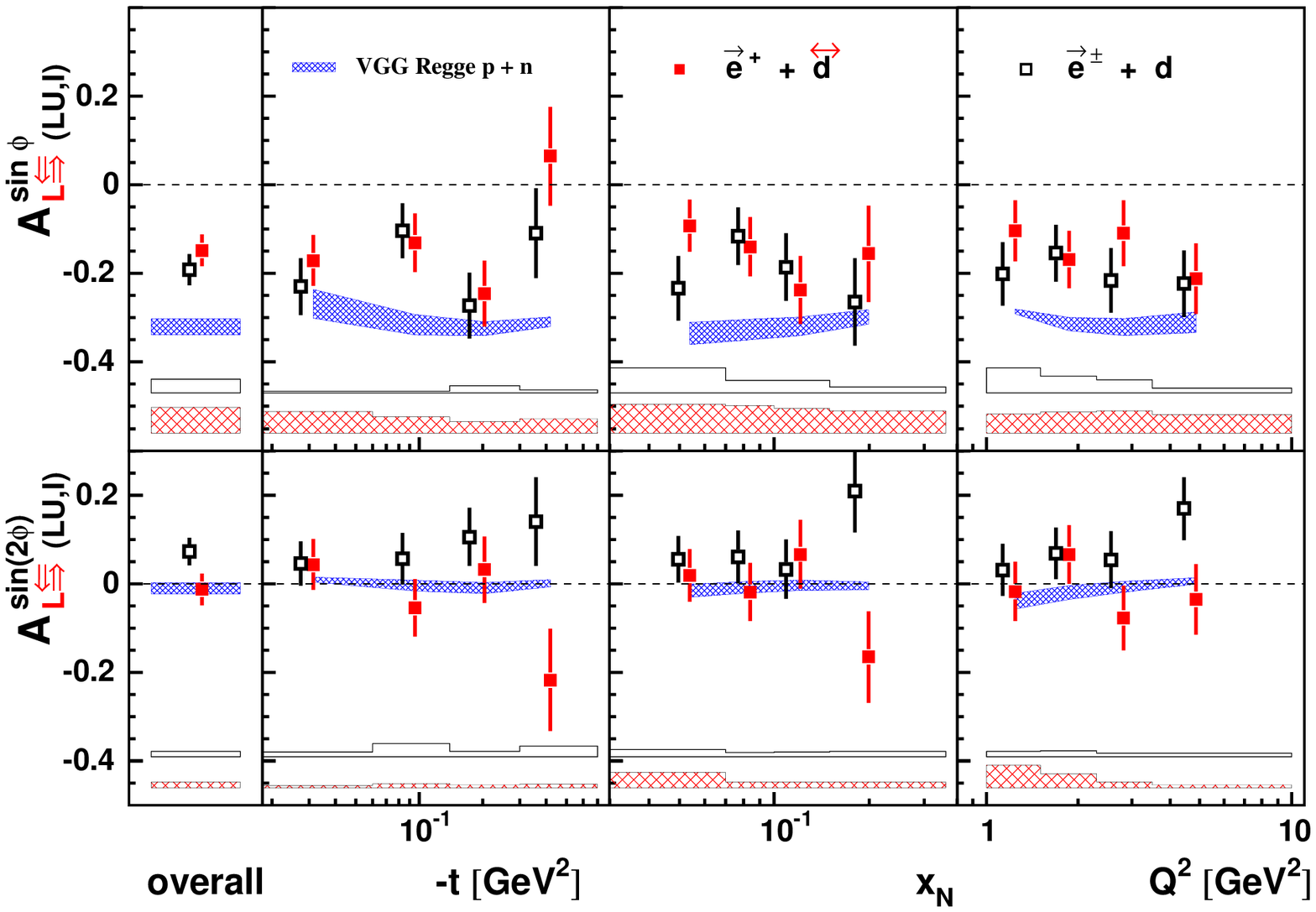}
\caption{Results from the present work (red filled squares) representing 
single-charge beam-helicity asymmetry amplitudes 
$A_{\chk{\mathrm{\scriptscriptstyle
L{\stackrel{\Leftarrow}{\scriptscriptstyle \Rightarrow}}}}}^{\sin(n\phi)}$ 
describing the dependence of the sum of squared DVCS and interference 
terms on the beam helicity, for a tensor polarization of $P_{zz}=0.827$ 
(indicated by the symbol \chk{$\leftrightarrow$}). The black open squares 
represent charge-difference amplitudes $\ALUI^{\sin(n\phi)}$ from only the 
interference term, extracted from unpolarized deuterium 
data~\cite{deuteron_unpol_draft}. The error bars represent the statistical 
uncertainties, while the coarsely hatched (open) bands represent the 
systematic uncertainties of the filled (open) squares. There is an 
additional overall 1.9$\%$ (2.4$\%$) scale uncertainty arising from the 
uncertainty in the measurement of the beam polarization in the case of 
polarized (unpolarized) deuterium data. The points for unpolarized 
deuterium data are slightly shifted to the left for better visibility. The 
finely hatched band shows the results of theoretical calculations for the 
combination of incoherent scattering on proton and neutron, using variants 
of the VGG double-distribution model~\cite{Vanderhaeghen:1999xj,Vdhcode} 
with a Regge ansatz for modeling the $t$ dependence of 
GPDs~\cite{Goeke:2001tz}.}
\label{fig:alpm}
\end{figure}

\begin{figure}[!tb]
\includegraphics[width=1\columnwidth]{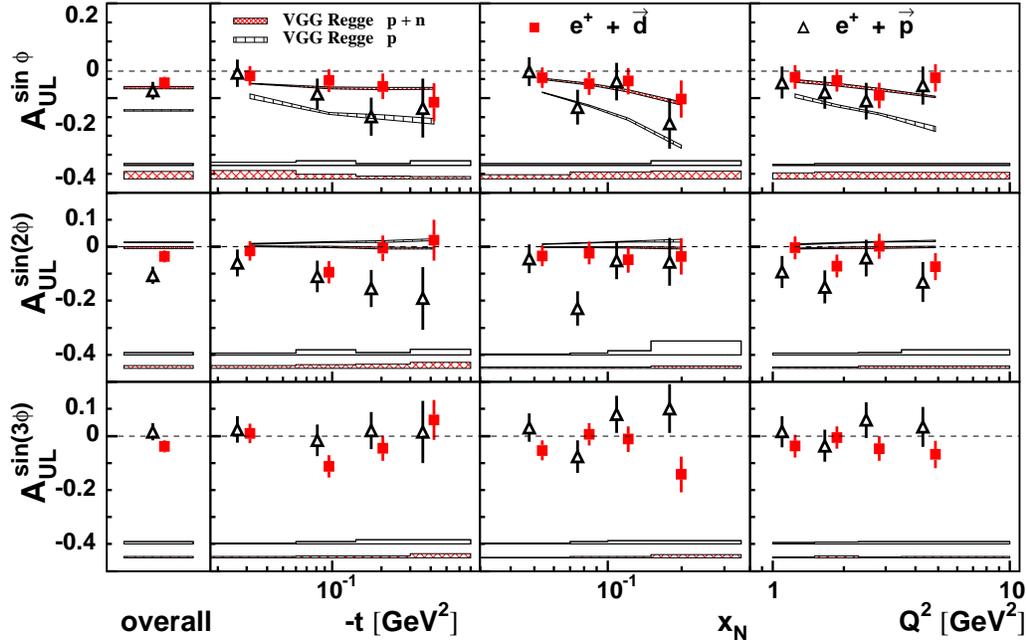}
\caption{Single-charge target-spin asymmetry amplitudes describing the 
dependence of the sum of squared DVCS and interference terms on the target 
vector polarization, for a tensor polarization of $P_{zz}=0.827$. The 
squares represent the results from the present work. The triangles denote 
the corresponding amplitudes extracted from longitudinally polarized 
hydrogen data~\cite{proton_pol_draft}. The error bars (bands) represent 
the statistical (systematic) uncertainties. The finely hatched bands have 
the same meaning as in Fig.~\ref{fig:alpm}. There is an additional overall 
4.0$\%$ (4.2$\%$) scale uncertainty arising from the uncertainty in the 
measurement of the target polarization in the case of deuterium 
(hydrogen). The points for hydrogen are slightly shifted to the left for 
better visibility.}
\label{fig:aul}
\end{figure}

\begin{figure}[!tb]
\includegraphics[width=1\columnwidth]{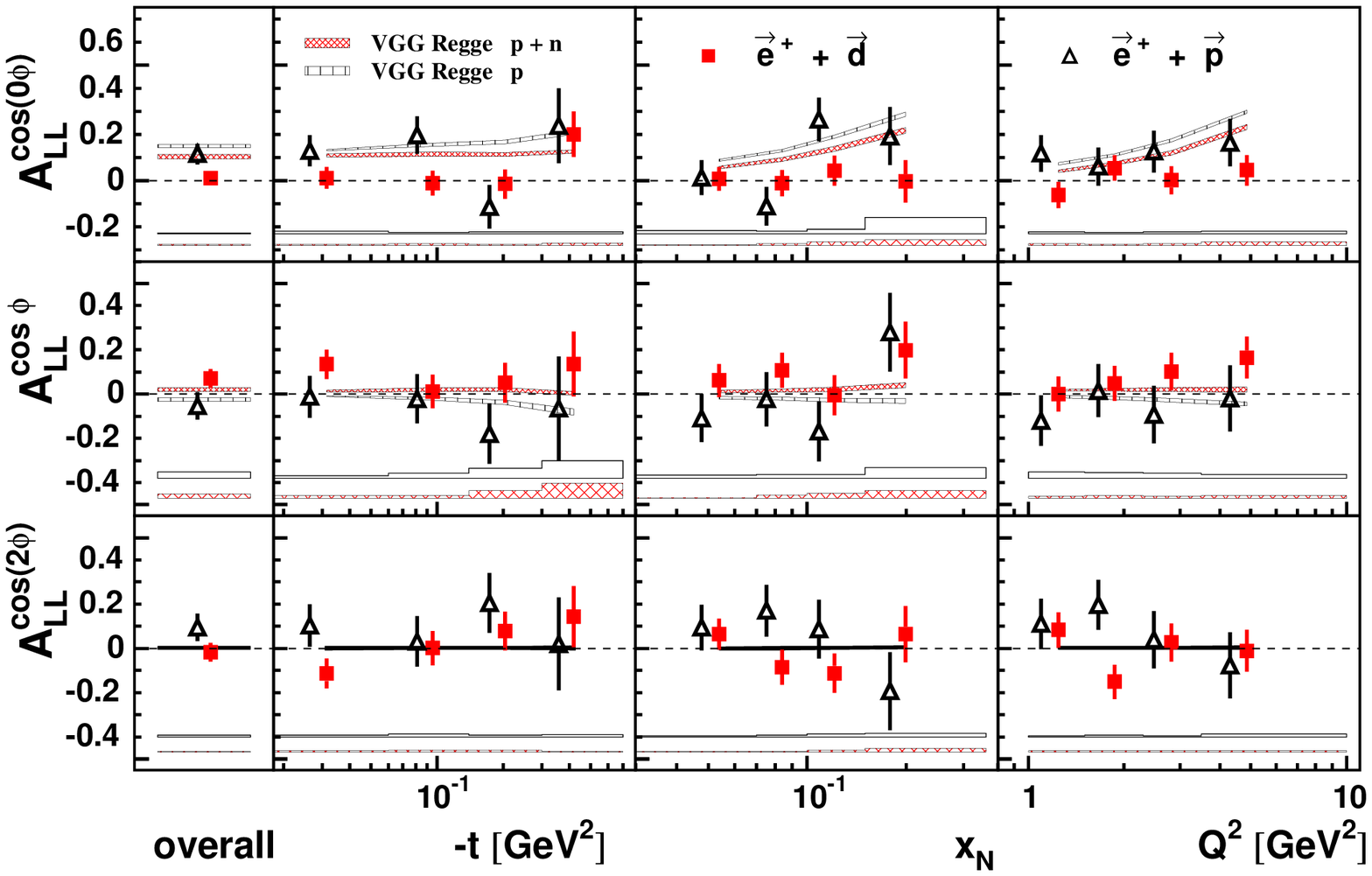}
\caption{Single-charge double-spin asymmetry amplitudes describing the 
dependence of the sum of Bethe-Heitler, squared DVCS and interference 
terms on the product of the beam helicity and target vector polarization, 
for a tensor polarization of $P_{zz}=0.827$. The plotted symbols and bands 
have the same meaning as in Fig.~\ref{fig:aul}. There is an additional 
overall 4.4$\%$ (5.3$\%$) scale uncertainty arising from the uncertainties 
in the measurement of the beam and target polarizations in the case of 
deuterium (hydrogen) data.}
\label{fig:all}
\end{figure}

\begin{figure}[!tb]
\includegraphics[width=1\columnwidth]{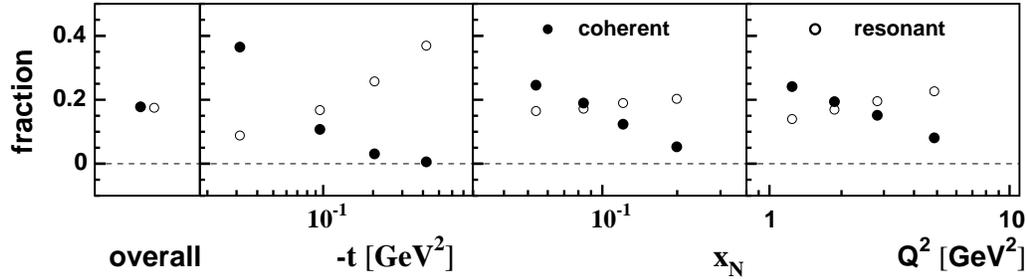}
\caption{Simulated yield fractions of coherent and resonant production}
\label{fig:fraction}
\end{figure}

The results for the Fourier amplitudes of the single-charge asymmetries 
$\CalALPM(e_\ell=+1,P_{zz},\phi)$, $\CalAUL(e_\ell=+1,P_{zz},\phi)$ and 
$\CalALL(e_\ell=+1,P_{zz},\phi)$ are presented in 
Figs.~\ref{fig:alpm}--\ref{fig:all} as a function of $-t$, $x_N$, or $Q^2$ 
and are also given in Table~\ref{tb:table2}. While the variable $x_D$ 
would be the appropriate choice to  present experimental results for pure 
coherent scattering, the nucleonic Bjorken variable $x_N$ is the practical 
choice in this case where incoherent scattering dominates over most of the 
kinematic range. The `overall' results in the left columns correspond to 
the entire HERMES kinematic acceptance. Figure~\ref{fig:alpm} shows the 
amplitudes $\ALPM^{\sin(n\phi)}$ related to beam helicity only, while 
Figs.~\ref{fig:aul} and \ref{fig:all} show the amplitudes 
$\AUL^{\sin(n\phi)}$, which relate to target vector polarization only, 
and the amplitudes $\ALL^{\cos(n\phi)}$, which relate to the product of 
beam helicity and target vector polarization. Table~\ref{tb:table3} and 
Fig.~\ref{fig:fraction} show in each kinematic bin the estimated 
fractional contributions to the yield from the coherent process and from 
processes leading to baryonic resonant final states. They are obtained 
from a Monte Carlo simulation using an exclusive-photon generator 
described in Ref.~\cite{deuteron_unpol_draft}.

The values for the $\sin \phi$ amplitude of the asymmetry $\CalALPM$ in 
Fig.~\ref{fig:alpm} are found to be significantly negative, while the 
$\sin (2\phi)$ amplitude is found to be consistent with zero. 
Figure~\ref{fig:alpm} also presents for comparison the amplitudes of the 
charge-difference asymmetry $\CalALUI$ extracted from a previous 
measurement on unpolarized deuterons~\cite{deuteron_unpol_draft}. Under the 
same approximations as those leading to Eq.~\ref{eq:Dpmtilde-short}, 
$\CalALPM$ is expected to differ from $\CalALUI$ (only if $P_{zz} \ne 0$) 
due only to a term involving the CFF ${\cal H}_5$. Figure~\ref{fig:alpm} 
shows that these two asymmetries are found to be consistent in most 
kinematic regions, except possibly for the last $-t$ or $x_N$ bin in the 
case of $\sin(2\phi)$. (The overall results differ by only 1.7 standard 
deviations in the total experimental uncertainties.\footnote{Here 
and hereafter we neglect any possible correlations arising from common 
treatments of different data sets.}) The consistency in the first $-t$ 
bin, where the contribution from coherent scattering is significant, 
suggests that there is no distinctive contribution from ${\cal H}_5$, as 
was observed in the case of the corresponding forward 
limit~\cite{hermes:b1,hermes:g1}.

In the first $-t$ bin, the asymmetry amplitude $\ALPMcoh^{\sin \phi}$ for 
pure coherent scattering on a polarized deuterium target was estimated 
from the measured asymmetry by correcting for the incoherent contributions 
of the proton and neutron and their resonances (see 
Ref.~\cite{deuteron_unpol_draft}). This correction is based on the 
assumption that for the incoherent contribution of the proton, 
$\ALPM^{\sin\phi}(P_{zz}=0.827) \approx \ALUI^{\sin\phi}$ where the latter 
was measured on a hydrogen target~\cite{proton_unpol_draft}. The 
fractional contributions and the asymmetry for incoherent scattering from 
the neutron was taken from the Monte Carlo calculation described in 
section~\ref{sec:background}, with uncertainties equal to their magnitude. 
The result for the asymmetry amplitude $\ALPMcoh^{\sin 
\phi}(P_{zz}=0.827)$ is estimated to be $-0.12 \pm 0.17(\rm {stat.}) \pm 
0.14(\rm {syst.}) \pm 0.02(\rm {model})$, where the systematic uncertainty 
is propagated from only the corresponding experimental uncertainties. 
Within the uncertainties there is no evidence of a difference between this 
value and the value for the asymmetry amplitude $\ALUcoh^{\sin\phi} = 
-0.29 \pm 0.18(\rm {stat.}) \pm 0.03(\rm {syst.})$ previously estimated 
for coherent scattering on an unpolarized deuterium target, using a 
disjoint HERMES data set for an unpolarized deuterium 
target~\cite{deuteron_unpol_draft}, but using the same data set for a 
hydrogen target.

\begin{table}
\caption{Simulated fractional contributions of coherent and resonant 
processes on a deuteron, in each kinematic bin.}
\small
\begin{center}
\begin{tabular}{c|ccccrr}
\hline\noalign{\smallskip}
\multicolumn{2}{c}{Kinematic bin} & $\langle -t \rangle$ & $\langle x_N 
\rangle$ & $\langle Q^2 \rangle$ & \hspace{0.25cm} Coherent & 
\hspace{0.25cm} Resonant\\
\multicolumn{2}{c}{} &[GeV$^2$] & &[GeV$^2$] & & \\
\noalign{\smallskip}
\hline\noalign{\smallskip}
\multicolumn{2}{c}{Overall} & 0.13 & 0.10 & 2.5 & 0.177 & 0.177 \\
\noalign{\smallskip}
\hline\noalign{\smallskip}
\multirow{4}{*}{\rotatebox{90}{\mbox{$-t$[GeV$^2$]}}}
& 0.00 - 0.06 & 0.03 & 0.08 & 1.9 & 0.364 & 0.088 \\
& 0.06 - 0.14 & 0.10 & 0.10 & 2.5 & 0.107 & 0.168 \\
& 0.14 - 0.30 & 0.20 & 0.11 & 2.9 & 0.030 & 0.257 \\
& 0.30 - 0.70 & 0.42 & 0.12 & 3.5 & 0.006 & 0.369 \\
\noalign{\smallskip}
\hline\noalign{\smallskip}
\multirow{4}{*}{\rotatebox{90}{\mbox{$x_N$}}}
& 0.03 - 0.07 & 0.11 & 0.05 & 1.4 & 0.246 & 0.164 \\
& 0.07 - 0.10 & 0.11 & 0.08 & 2.1 & 0.189 & 0.172 \\
& 0.10 - 0.15 & 0.14 & 0.12 & 3.1 & 0.123 & 0.189 \\
& 0.15 - 0.35 & 0.20 & 0.20 & 5.0 & 0.053 & 0.202 \\
\noalign{\smallskip}
\hline\noalign{\smallskip}
\multirow{4}{*}{\rotatebox{90}{\mbox{$Q^{2}$[GeV$^2$]}}}
& 1.0 - 1.5 & 0.09 & 0.06 & 1.2 & 0.241 & 0.139 \\
& 1.5 - 2.3 & 0.11 & 0.08 & 1.9 & 0.194 & 0.169 \\
& 2.3 - 3.5 & 0.14 & 0.11 & 2.8 & 0.151 & 0.196 \\
& 3.5 - 10.0 & 0.20 & 0.17 & 4.9 & 0.080 & 0.226 \\
\noalign{\smallskip}
\hline\noalign{\smallskip}
\end{tabular}
\label{tb:table3}
\end{center}
\end{table}
The extracted values for the $\sin \phi$ and $\sin (2\phi)$ amplitudes of 
the single-charge asymmetry $\CalAUL$ measured on a longitudinally 
polarized deuterium target are shown in Fig.~\ref{fig:aul}. The `overall' 
values are slightly negative by less than 1.5 standard deviations of the 
total experimental uncertainty. For coherent scattering on the deuteron, 
the amplitude $\AUL^{\sin\phi}$ is sensitive to the imaginary part of a 
combination of deuteron CFFs $\widetilde{\cal H}_1$, ${\cal H}_1$ and 
${\cal H}_5$ weighted with the elastic form factors of the deuteron $G_1$ 
and $G_2$ (see Eq.~\ref{App-AUL-short}). In particular, for the first $-t$ 
bin where $\langle x_D \rangle = 0.04$, $G_1$ is about 30 times larger 
than $\frac{x_D}{2} G_2$. Thus the CFF $\widetilde{\cal H}_1$ may 
influence the resulting $\AUL^{\sin\phi}$ amplitude in the first $-t$ bin 
where the coherent process contributes approximately 40$\%$. For 
comparison, the same amplitudes measured on a longitudinally polarized 
hydrogen target~\cite{proton_pol_draft} are also shown in 
Fig.~\ref{fig:aul}. The $\sin \phi$ amplitude shows consistency between 
deuterium and hydrogen data both for the `overall' result and the 
kinematic projections on $-t$, $x_N$, and $Q^2$. In this comparison, no 
account was taken of the 7.5$\%$ depolarization of nucleons in the 
deuteron due to the 5$\%$ admixture of the $D$-state~\cite{Umnikov}. The 
`overall' results on the $\sin (2\phi)$ amplitude differ between 
the two targets  by 1.5 standard deviations of the total experimental 
uncertainties, mainly due to the region of large $-t$, but in only one 
$x_N$ bin. The `overall' result on the asymmetry amplitude 
$\AUL^{\sin(3\phi)}$ is slightly negative by less than 1.7 standard 
deviations of the total experimental uncertainty. The $\sin (3\phi)$ 
amplitude shows consistency between deuterium and hydrogen data, 
accounting for the total experimental uncertainties of the corresponding 
measurements, except possibly for the highest $x_N$ bin.

The $\ALL^{\cos (n\phi)}$ amplitudes of the single-charge double-spin 
asymmetry measured using longitudinally polarized deuteron data and 
presented in Fig.~\ref{fig:all} are found to be compatible with zero, 
although the $\ALL^{\cos \phi}$ amplitude is positive by 1.6 standard 
deviations of the total experimental uncertainty. Within the 
uncertainties, these asymmetry amplitudes do not show significant 
differences from those measured on a longitudinally polarized hydrogen 
target~\cite{proton_pol_draft}, except possibly for the overall result for 
the amplitude $\ALL^{\cos (0\phi)}$, where there is observed a discrepancy 
of 1.9 standard deviations in the total experimental uncertainties.

The finely hatched bands in Figs.~\ref{fig:alpm}--\ref{fig:all} 
represent results of theoretical calculations based on the GPD model 
described in Ref.~\cite{Vanderhaeghen:1999xj}, using the VGG computer 
program of Ref.~\cite{Vdhcode}. The Regge ansatz for modeling the $t$ 
dependence of GPDs~\cite{Goeke:2001tz} is used in these calculations. 
The model~\cite{Vanderhaeghen:1999xj} is an implementation  of the 
double-distribution concept~\cite{Mul94,Rad97} where the kernel of the 
double distribution contains a profile function that determines the 
dependence on $\xi$, controlled by a parameter $b$~\cite{Musatov} for each 
quark flavor. The cross sections are calculated as the sum of the 
incoherent processes on the proton and neutron in each kinematic bin. (No 
computer program is available simulating coherent scattering on the 
deuteron.) The width of the theoretical bands in 
Figs.~\ref{fig:alpm}--\ref{fig:all} corresponds to the range of values of 
the asymmetry amplitudes obtained by varying the profile parameters 
$b_{\rm val}$ and $b_{\rm sea}$ between unity and infinity. In the 
comparison of these predictions with experimental results, it should be 
noted that the effect of the $D$-state of the deuteron on the polarization 
of the nucleons inside the deuteron was not taken into account.

The model calculations predict a magnitude of the $\sin \phi$ harmonic of 
the single-charge beam-helicity asymmetry that exceeds that of the data by 
about a factor of two, a situation similar to that found in the case of a 
hydrogen target~\cite{proton_unpol_draft}. On the other hand the 
predictions are in good agreement with data for single-charge target-spin 
asymmetries. A large difference appears between the predictions for the 
$\sin \phi$ harmonic of this asymmetry on the deuteron and proton targets, 
arising entirely from the contributions of the neutron. The data are 
consistent with this difference, but lack the precision to confirm the 
large positive prediction of the neutron asymmetry by this model. The 
predictions are in good agreement with the single-charge double-spin 
asymmetry amplitudes, aside from the $\cos (0\phi)$ harmonic. Here the 
theoretical predictions for both the deuteron and proton, which are 
dominated by the BH contribution, are significantly positive, in agreement 
with the proton data, while the more precise deuteron data are consistent 
with zero. The small contribution of coherent scattering to the overall 
result, with a predicted negative asymmetry~\cite{theor_deu}, is expected 
to slightly reduce this asymmetry amplitude for the deuteron.

\subsection{The beam-charge, charge-averaged, and 
beam-charge\,$\otimes$target-spin asymmetries}
\label{beam_charge_asymm}
\begin{figure}[!hbp]
\includegraphics[width=1\columnwidth]{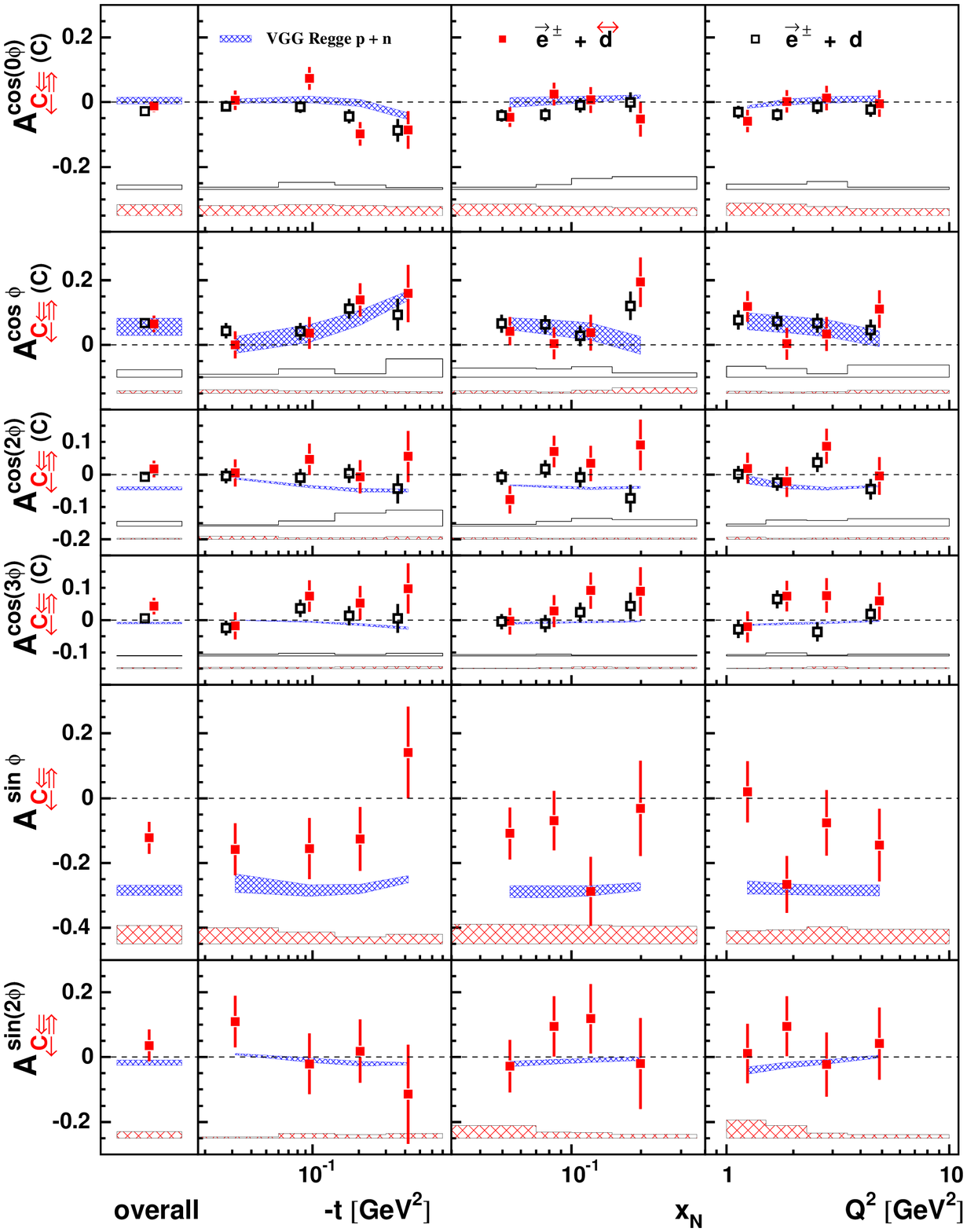}
\caption{Results from the present work (red filled squares) representing 
single-beam-helicity charge asymmetry amplitudes 
$A_{\chk{\mathrm{\scriptscriptstyle
C{\stackrel{\Leftarrow}{\scriptscriptstyle \Rightarrow}}}}}^{\cos(n\phi)}$
and $A_{\chk{\mathrm{\scriptscriptstyle
C{\stackrel{\Leftarrow}{\scriptscriptstyle \Rightarrow}}}}}^{\sin(n\phi)}$, 
for $P_\ell=-0.530$ and a tensor polarization of $P_{zz}=0.827$ 
(indicated by the symbol \chk{$\leftrightarrow$}). The black open squares 
are $\AC^{\cos(n\phi)}$ amplitudes extracted from data recorded with an 
unpolarized beam and unpolarized deuterium 
target~\cite{deuteron_unpol_draft}. The error bars and bands and 
finely hatched bands have the same meaning as in Fig.~\ref{fig:alpm}. The 
points for unpolarized deuterium data are slightly shifted to the left for 
better visibility. There is an additional overall 2.2$\%$ scale 
uncertainty for the $\ACPMM^{\sin(n\phi)}$ amplitudes arising from the 
uncertainty in the measurement of the beam polarization.}
\label{fig:acpm}
\end{figure}

\begin{figure}[ht]
\includegraphics[width=1\columnwidth]{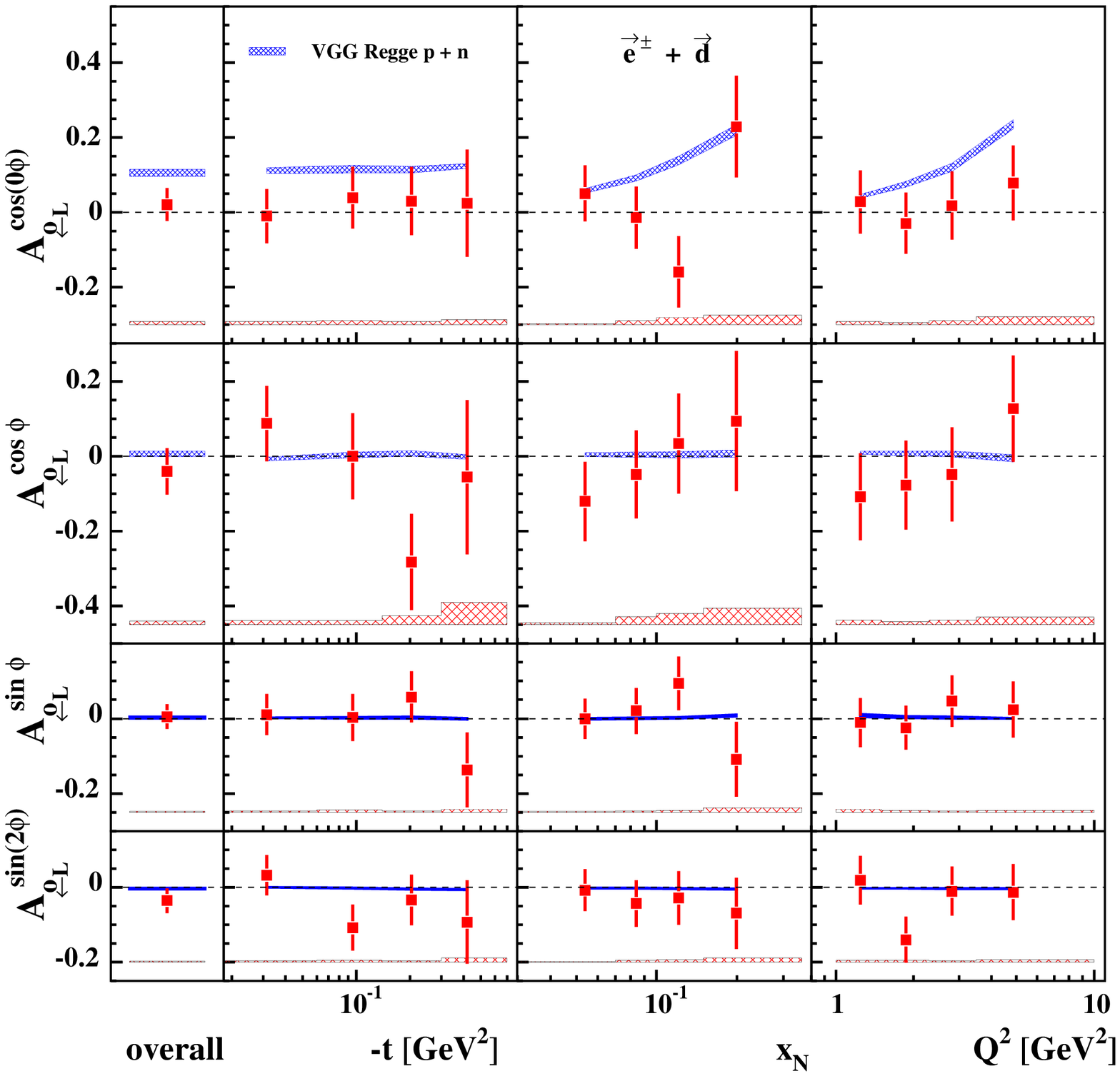}
\caption{Kinematic dependence of the charge-averaged single-beam-helicity 
target-spin asymmetry amplitudes $\AP^{\cos(n\phi)}$ and 
$\AP^{\sin(n\phi)}$, for $P_\ell=-0.530$ and a tensor polarization of 
$P_{zz}=0.827$. The plotted symbols and bands have the same meaning as in 
Fig.~\ref{fig:acpm}. There is an additional overall 5.3$\%$ (5.7$\%$) 
scale uncertainty for the extracted $\AP^{\sin(n\phi)}$ 
$\left(\AP^{\cos(n\phi)}\right)$ amplitudes arising from the uncertainties 
in the measurement of the target (beam and target) polarizations.}
\label{fig:ap}
\end{figure}

\begin{figure}[!tb]
\includegraphics[width=1\columnwidth]{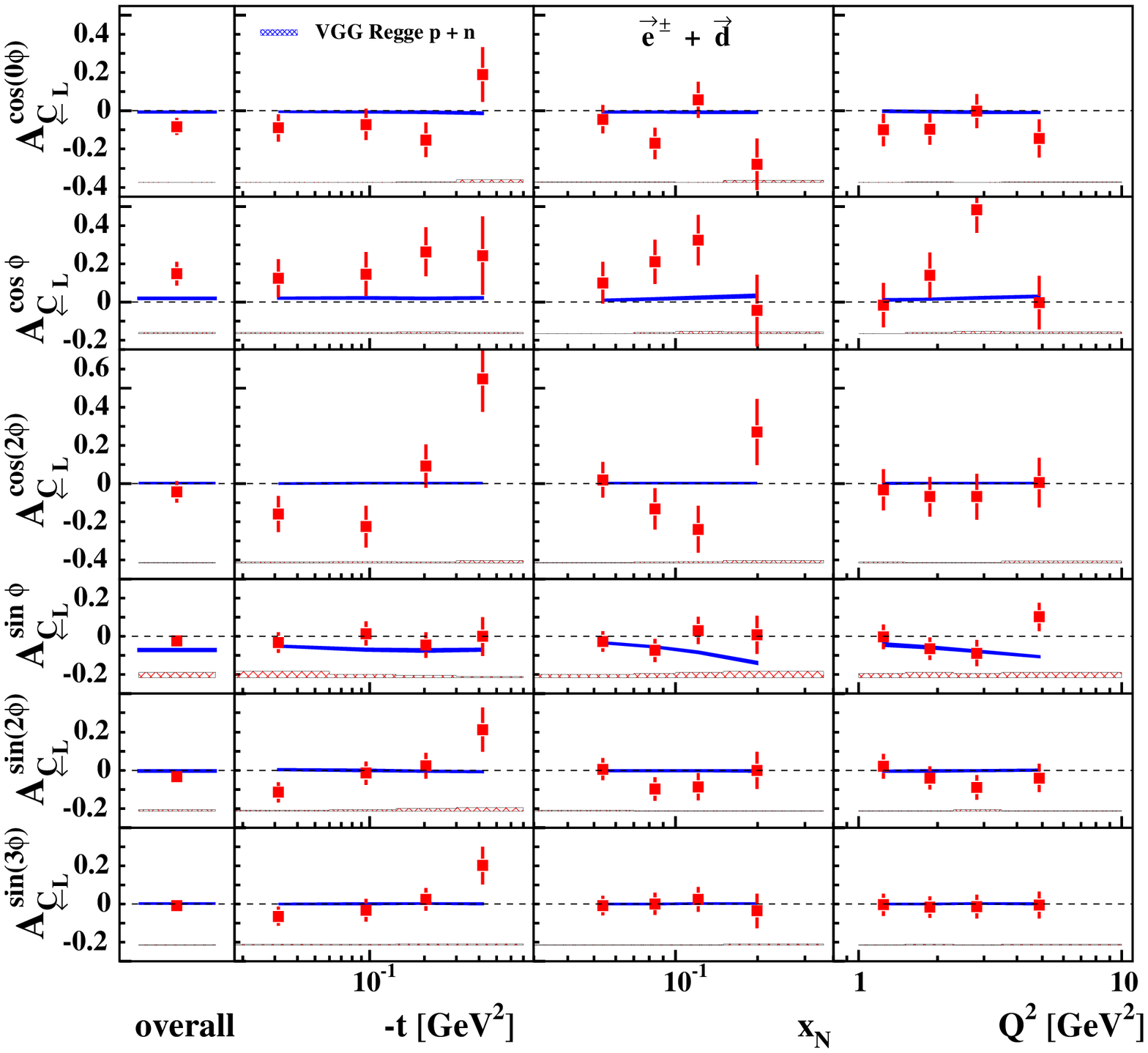}
\caption{Kinematic dependence of the single-beam-helicity 
beam-charge$\otimes$target-spin asymmetry amplitudes $\ACP^{\cos(n\phi)}$ and 
$\ACP^{\sin(n\phi)}$, for $P_\ell=-0.530$ and a tensor polarization of 
$P_{zz}=0.827$. The plotted symbols and bands have the same meaning as in 
Fig.~\ref{fig:acpm}. There is an additional overall 5.3$\%$ (5.7$\%$) 
scale uncertainty for the extracted $\ACP^{\sin(n\phi)}$ 
$\left(\ACP^{\cos(n\phi)}\right)$ amplitudes arising from the 
uncertainties in the measurement of the target (beam and target) 
polarizations.} 
\label{fig:acp}
\end{figure}

The results for the Fourier amplitudes of the single-beam-helicity
asymmetries are presented in Figs.~\ref{fig:acpm}--\ref{fig:acp}. More 
specifically, Figs.~\ref{fig:acpm}, \ref{fig:ap}, and \ref{fig:acp} show 
the $\cos(n\phi)$ and $\sin(n\phi$) harmonics of the asymmetry 
$\CalACPM(P_\ell,P_{zz},\phi)$, $\CalAP(P_\ell,P_{zz},\phi)$ and 
$\CalACP(P_\ell,P_{zz},\phi)$, respectively (see also 
Tables~\ref{tb:table4}--\ref{tb:table6}), for $P_\ell=-0.530 \pm 0.012$ 
and  $P_{zz}=0.827 \pm 0.027$.

The only overall results for the asymmetry $\CalACPM$ in 
Fig.~\ref{fig:acpm} that are found to be significantly non-zero are the 
$\cos \phi$ and $\sin \phi$ amplitudes. The theoretical calculations for 
incoherent scattering predict that the results for the amplitudes 
$\ACPMM^{\cos(n\phi)}$ should strongly resemble those for the amplitudes 
$\AC^{\cos(n\phi)}$ measured with an unpolarized beam on an unpolarized 
deuterium target \cite{deuteron_unpol_draft}. The data confirm this 
resemblance, even in the first $-t$ bin where coherent scattering 
contributes about 40$\%$ of the yield. This is another indication that the 
CFF ${\cal H}_5$~\cite{theor_deu}, in this case its real part, makes no 
distinctive contribution to coherent scattering off deuterons, similar to 
the case of $\ALPM^{\sin \phi}$, as was noted in the discussion in 
Section~\ref{single_double_asymm} about the dependence of $\ALPM^{\sin 
\phi}$ of Eq.~\ref{eq:Dpmtilde-short} on the imaginary part of this CFF.

The numerators of the $\ACPMM^{\sin(n\phi)}$ amplitudes shown in 
Fig.~\ref{fig:acpm} differ from those of the $\sin(n\phi)$ amplitudes of 
the $\CalALPM$ asymmetry shown in Fig.~\ref{fig:alpm} only by squared DVCS 
terms. Furthermore, the cross sections $\rd\CLCPM$ and $\rd\CULPM$ in the 
denominators of these two asymmetries should be similar because they are 
dominated by Bethe-Heitler contributions. Hence, these asymmetry 
amplitudes are expected to be similar, and within the statistical accuracy 
this is indeed found to be the case.

The $\cos (n\phi)$ amplitudes of the asymmetry $\CalAP$ in 
Fig.~\ref{fig:ap} contain a sum of BH and squared DVCS even harmonics, and 
relate to the longitudinal vector polarization of the target. However, 
even where the BH contribution dominates the numerator of the asymmetry 
amplitude $\AP^{\cos (0\phi)}$ for incoherent scattering at not small 
$-t$, the data are found to be consistent with zero, and differing by 1.7 
standard deviations in the total experimental uncertainty from the 
positive prediction for the overall result. The $\sin(n\phi)$ amplitudes 
of the asymmetry $\CalAP$ in Fig.~\ref{fig:ap} receive contributions from 
the pure squared DVCS harmonics only, and are found to be consistent with 
zero.

Of particular interest are the $\ACP^{\cos (n\phi)}$ and $\ACP^{\sin 
(n\phi)}$ amplitudes shown in Fig.~\ref{fig:acp}, which represent 
respectively the even and odd vector-polarization related harmonics of the 
interference term only, receiving no contribution from pure BH and DVCS 
terms. The theoretical predictions for the $\cos (n\phi)$ harmonics are 
negligibly small, while the data differ from zero by about two standard 
deviations for the first two harmonics. As expected and observed in the 
case of unpolarized hydrogen and deuterium targets, the $\cos (0\phi)$ and 
$\cos \phi$ harmonics are found to have opposite signs.

Like the asymmetry amplitude $\AUL^{\sin(\phi)}$, in the first $-t$ bin 
the asymmetry  amplitude $\ACP^{\sin \phi}$ is sensitive to the imaginary 
part of the deuteron CFF $\widetilde{\cal H}_1$. Within their statistical 
accuracies, they are found to be consistent, although $\AUL^{\sin \phi}$ 
receives also a contribution from the squared DVCS term (see 
Eq.~\ref{eq:sigma00_assaul}). The asymmetry amplitude $\ACP^{\cos \phi}$ 
is sensitive to the real part of the deuteron CFF $\widetilde{\cal H}_1$. 
Unlike the corresponding harmonic $\ALL^{\cos \phi}$, it does not receive 
a contribution from the Bethe--Heitler term. The $\sin (n\phi)$ harmonics 
are found to be consistent with zero and also with the small negative 
prediction in the case of the $\sin \phi$ harmonic.

From the definitions of the asymmetries $\CalAUL$, $\CalALL$, $\CalAP$ and 
$\CalACP$ in Eqs.~\ref{eq:ulphi}, \ref{eq:sigma00_assall_beg}, 
\ref{eq:ap_beg}, and \ref{eq:acp_beg}, and also from examination of Table 
2, it can be seen that they are related. In the case of approximate 
equality of $\rd\CLCPM$ and $\rd\CULPM$, the following relations hold 
between the asymmetry amplitudes:
\begin{eqnarray}
\AUL^{\sin(n\phi)}\simeq \AP^{\sin(n\phi)}+\ACP^{\sin(n\phi)} \,, \, \, \, \, 
n = 1, 2 \, , \label{eq:corr1}\\
\ALL^{\cos(n\phi)}\simeq \AP^{\cos(n\phi)}+\ACP^{\cos(n\phi)} \,, \, \, \, \, 
n = 0, 1 \, .
\label{eq:corr2}
\end{eqnarray}
For most of the kinematic points, the differences between left and right 
hand sides of Eqs. 45 and 46 are found below 1.2 standard deviations of 
the total experimental uncertainties, while for the remaining six points 
they are between 1.5 and 2.0. Note that here the correlations between two 
asymmetries from the right hand sides are taken into account.

\subsection{The beam-helicity\,$\otimes$tensor asymmetry $\CalALzz$}
\label{subsec:alzz}
The definition of the asymmetry $\CalALzz$ is given in 
Eq.~\ref{equ:tensor}. As mentioned in Section~\ref{sec:experiment}, for 
the extraction of this asymmetry, the data taken with a positron beam and 
with the average target tensor polarization $P_{zz} = -1.656$ are used in 
combination with the positron data collected on a longitudinally polarized 
deuterium target with $P_{zz} = 0.827$. The same maximum likelihood 
technique~\cite{hermes_ttsa} unbinned in azimuthal angle $\phi$ was used 
to extract the $\CalALzz^{\sin(n\phi)}$ Fourier amplitudes. The 
$\CalALzz^{\sin \phi}$ amplitude is found to have the value 
$\CalALzz^{\sin \phi} = -0.130 \pm 0.121(stat.) \pm 0.051(syst.)$ when 
extracted in the entire kinematic range of the data set, while in the 
region $-t < 0.06 \rm\,GeV^2$, where the contribution from coherent 
scattering on a longitudinally polarized deuteron is approximately 40$\%$, 
this value is found to be $0.074 \pm 0.196(stat.) \pm 0.022(syst.)$. 
These results are subject to an additional scale uncertainty of 3.8$\%$ 
arising from beam and target-tensor polarizations. The Fourier amplitudes 
related to higher twist are found to be compatible with zero within the 
statistical uncertainties. This `zero' result for the 
beam-helicity$\otimes$tensor asymmetry $\CalALzz$ extracted independently 
of the results for $\CalALUI$ and $\CalALPM$ confirms that there 
is no distinctive contribution from the deuteron CFF ${\cal H}_5$ for 
coherent scattering.

\section{Summary}
\label{sec:summary}
Azimuthal asymmetries with respect to target polarization alone and also 
combined with beam helicity and/or beam charge for hard exclusive 
electroproduction of real photons  in deep-inelastic scattering from a 
longitudinally polarized deuterium target are measured for the first time. 
The asymmetries are attributed to the interference between the deeply 
virtual Compton scattering and Bethe--Heitler processes. The asymmetries 
are observed in the exclusive region $-(1.5)^2{\rm\,GeV}^2 < M_X^2 
<(1.7)^2{\rm\,GeV}^2$ of the squared missing mass. The dependences of 
these asymmetries on $-t$, $x_N$, or $Q^2$ are investigated. The results 
include the coherent process $e \, d \to e \, d \, \gamma$ and the 
incoherent process $e \, d \to e \, p \, n \, \gamma$ where in addition a 
nucleon may be excited to a resonance. Within the total experimental 
uncertainties, the results of the sinusoidal (cosinusoidal) amplitudes of 
the asymmetry $\CalALPM$ ($\CalACPM$) extracted from a data set with 
$P_{zz} =0.827$ (corresponding to a small population for the $\Lambda =0$ 
state) resemble those for the amplitudes extracted from unpolarized 
deuterium data at HERMES. Therefore, no indication of effects of tensor 
polarization was found at small values of $-t$, in particular in the 
region  $-t < 0.06 \rm\,GeV^2$ where the coherent process contributes up 
to 40$\%$. Neither the $\CalAUL^{\sin(n\phi)}$ nor $\CalALL^{\cos(n\phi)}$ 
amplitudes measured on longitudinally polarized deuterons show significant 
differences compared with those extracted from longitudinally polarized 
protons, considering the total experimental uncertainties. (Statistically 
marginal differences are observed for $\CalAUL^{\sin(2\phi)}$ and 
$\CalALL^{\cos(0\phi)}$).

The sinusoidal amplitudes of the tensor asymmetry $\ALzz$ are compatible 
with zero for the whole kinematic range as well as for the region $-t < 
0.06 \rm\,GeV^2$ within the accuracy of the measurement. This suggests 
that differences between the leading amplitudes of the asymmetries 
$\CalALUI$ and $\CalALPM$ for coherent scattering from unpolarized and 
longitudinally polarized deuterium targets, respectively, should be small. 
Indeed, within the total experimental uncertainties, no difference is seen 
between the reconstructed values of the asymmetry amplitudes 
$\ALPMcoh^{\sin\phi}$ and $\ALUcoh^{\sin\phi}$.

In conclusion, even in the region $-t < 0.06 \rm\,GeV^2$ where the 
coherent process contributes about 40$\%$, all asymmetries on deuterium 
that have (approximate) counterparts for hydrogen are found to be 
compatible with them. The data are unable to reveal any evidence of the 
influence of the Compton form factor ${\cal H}_5$ or features of the 
deuteron Compton form factors ${\cal H}_1$ and $\widetilde{\cal H}_1$ that 
distinguish them from the counterparts for the proton. Hence, coherent 
scattering presents no obvious signature in these data. The deuteron 
Compton form factor ${\cal H}_1$  appears to have a similar behavior as 
${\cal H}$ of the proton. The data were compared with theoretical 
calculations for only incoherent scattering, based on a well-known GPD 
model. Those asymmetries that are expected to resemble counterparts for a 
hydrogen target reveal the same shortcomings of the model calculations 
that appeared in comparisons with the hydrogen data.

\section{Acknowledgments}
\label{sec:Acknowledgments}
We gratefully acknowledge the DESY management for its support and the 
staff at DESY and the collaborating institutions for their significant 
effort. This work was supported by the Ministry of Economy and the 
Ministry of Education and Science of Armenia; the FWO-Flanders and IWT, 
Belgium; the Natural Sciences and Engineering Research Council of Canada; 
the National Natural Science Foundation of China; the Alexander von 
Humboldt Stiftung, the German Bundesministerium f\"ur Bildung und 
Forschung (BMBF), and the Deutsche Forschungsgemeinschaft (DFG); the 
Italian Istituto Nazionale di Fisica Nucleare (INFN); the MEXT, JSPS, and 
G-COE of Japan; the Dutch Foundation for Fundamenteel Onderzoek der 
Materie (FOM); the Russian Academy of Science and the Russian Federal 
Agency for Science and Innovations; the U.K.~Engineering and Physical 
Sciences Research Council, the Science and Technology Facilities Council, 
and the Scottish Universities Physics Alliance; the U.S.~Department of 
Energy (DOE) and the National Science Foundation (NSF); and the European 
Community Research Infrastructure Integrating Activity under the FP7 
"Study of strongly interacting matter (HadronPhysics2, Grant Agreement 
number 227431)".

\clearpage
\begin{table}
\caption{Results for azimuthal Fourier amplitudes of the single-charge 
asymmetries $\CalALPM$, $\CalAUL$ and $\CalALL$, extracted from 
longitudinally polarized deuteron data, for a tensor polarization of 
$P_{zz}=0.827$. Not included are the 1.9$\%$, 4.0$\%$ and 4.4$\%$ scale 
uncertainties for corresponding asymmetry amplitudes arising from the 
uncertainties in the measurement of the beam, target, beam and target 
polarizations, respectively.}
\tiny
\begin{center}
\begin{tabular}{r|ccccrrrr}
\noalign{\smallskip}
\hline\noalign{\smallskip}
\multicolumn{2}{c}{Kinematic bin}
&$\langle -t \rangle$ &$\langle x_N \rangle$ &$\langle Q^2
\rangle$ &$\ALPM^{\sin \, \phi}$\hspace{0.9cm}
&$\ALPM^{\sin \, (2\phi)}$\hspace{0.9cm}
&$\AUL^{\sin \, \phi}$\hspace{0.9cm}
&$\AUL^{\sin \, (2\phi)}$\hspace{0.9cm} \\
\multicolumn{2}{c}{} &[GeV$^2$] & &[GeV$^2$] &$\pm \rm \delta_{stat} \pm
\rm \delta_{syst}$ &$\pm \rm \delta_{stat} \pm \rm \delta_{syst}$ &$\pm
\rm \delta_{stat} \pm \rm \delta_{syst}$
&$\pm \rm \delta_{stat} \pm \rm \delta_{syst}$ \\
\noalign{\smallskip}
\hline\noalign{\smallskip}
\multicolumn{2}{c}{Overall} & 0.13 & 0.10 & 2.5 & $-0.148\pm0.036
\pm0.058$ & $-0.012\pm0.035\pm0.013$ & $-0.044\pm0.023\pm0.029$
& $-0.037\pm0.022\pm0.010$\\
\noalign{\smallskip}
\hline\noalign{\smallskip}
\multirow{4}{*}{\rotatebox{90}{\mbox{$-t$[GeV$^2$]}}}
&0.00-0.06 & 0.03 & 0.08 & 1.9 & $-0.171\pm0.058\pm0.049$
& $0.043\pm0.057\pm0.005$ & $-0.018\pm0.037\pm0.031$
& $-0.015\pm0.036\pm0.010$\\
&0.06-0.14 & 0.10 & 0.10 & 2.5 & $-0.131\pm0.066\pm0.037$
& $-0.053\pm0.065\pm0.010$ & $-0.036\pm0.042\pm0.018$
& $-0.094\pm0.041\pm0.013$\\
&0.14-0.30 & 0.20 & 0.11 & 2.9 & $-0.246\pm0.074\pm0.025$
& $0.032\pm0.075\pm0.007$ & $-0.057\pm0.047\pm0.012$
& $-0.006\pm0.048\pm0.015$\\
&0.30-0.70 & 0.42 & 0.12 & 3.5 & $0.064\pm0.111\pm0.032$
& $-0.217\pm0.115\pm0.008$ & $-0.116\pm0.071\pm0.009$
& $0.024\pm0.075\pm0.023$\\
\noalign{\smallskip}
\hline\noalign{\smallskip}
\multirow{4}{*}{\rotatebox{90}{\mbox{$x_N$}}}
&0.03-0.07 & 0.11 & 0.05 & 1.4 & $-0.093\pm0.058\pm0.064$
& $0.018\pm0.060\pm0.035$ & $-0.025\pm0.037\pm0.016$
& $-0.034\pm0.038\pm0.005$\\
&0.07-0.10 & 0.11 & 0.08 & 2.1 & $-0.140\pm0.067\pm0.062$
& $-0.019\pm0.066\pm0.013$ & $-0.046\pm0.042\pm0.026$
& $-0.023\pm0.042\pm0.006$\\
&0.10-0.15 & 0.14 & 0.12 & 3.1 & $-0.238\pm0.077\pm0.055$
& $0.066\pm0.077\pm0.014$ & $-0.037\pm0.049\pm0.026$
& $-0.048\pm0.049\pm0.006$\\
&0.15-0.35 & 0.20 & 0.20 & 5.0 & $-0.156\pm0.109\pm0.049$
& $-0.165\pm0.103\pm0.013$ & $-0.104\pm0.069\pm0.030$
& $-0.036\pm0.068\pm0.009$\\
\noalign{\smallskip}
\hline\noalign{\smallskip}
\multirow{4}{*}{\rotatebox{90}{\mbox{$Q^{2}$[GeV$^2$]}}}
&1.0-1.5 & 0.09 & 0.06 & 1.2 & $-0.103\pm0.068\pm0.043$
& $-0.017\pm0.067\pm0.051$ & $-0.022\pm0.043\pm0.023$
& $-0.004\pm0.042\pm0.005$\\
&1.5-2.3 & 0.11 & 0.08 & 1.9 & $-0.169\pm0.065\pm0.047$
& $0.065\pm0.066\pm0.032$ & $-0.035\pm0.041\pm0.026$
& $-0.071\pm0.042\pm0.006$\\
&2.3-3.5 & 0.14 & 0.11 & 2.8 & $-0.110\pm0.074\pm0.050$
& $-0.077\pm0.073\pm0.014$ & $-0.091\pm0.047\pm0.026$
& $-0.002\pm0.046\pm0.008$\\
&3.5-10.0 & 0.20 & 0.17 & 4.9 & $-0.212\pm0.079\pm0.042$
& $-0.036\pm0.080\pm0.006$ & $-0.025\pm0.050\pm0.024$
& $-0.073\pm0.050\pm0.008$\\
\noalign{\smallskip}
\hline\noalign{\smallskip}
\end{tabular}
\end{center}

\begin{center}
\begin{tabular}{r|ccccrrrr}
\hline\noalign{\smallskip}
\multicolumn{2}{c}{Kinematic bin}
&$\langle -t \rangle$ &$\langle x_N \rangle$ &$\langle Q^2
\rangle$ &$\AUL^{\sin \, (3\phi)}$\hspace{0.9cm}
&$\ALL^{\cos \, (0\phi)}$\hspace{0.9cm}
&$\ALL^{\cos \, \phi}$\hspace{0.9cm}
&$\ALL^{\cos \, (2\phi)}$\hspace{0.9cm} \\
\multicolumn{2}{c}{} &[GeV$^2$] & &[GeV$^2$]
&$\pm \rm \delta_{stat} \pm \rm \delta_{syst}$
&$\pm \rm \delta_{stat} \pm \rm \delta_{syst}$
&$\pm \rm \delta_{stat} \pm \rm \delta_{syst}$
&$\pm \rm \delta_{stat} \pm \rm \delta_{syst}$ \\
\noalign{\smallskip}
\hline\noalign{\smallskip}
\multicolumn{2}{c}{Overall} & 0.13 & 0.10 & 2.5
& $-0.039\pm0.022\pm0.004$ & $0.011\pm0.029\pm0.004$
& $0.072\pm0.042\pm0.019$ & $-0.017\pm0.042\pm0.005$\\
\noalign{\smallskip}
\hline\noalign{\smallskip}
\multirow{4}{*}{\rotatebox{90}{\mbox{$-t$[GeV$^2$]}}}
&0.00-0.06 & 0.03 & 0.08 & 1.9
& $0.009\pm0.036\pm0.005$ & $0.012\pm0.048\pm0.005$
& $0.136\pm0.066\pm0.010$ & $-0.115\pm0.068\pm0.008$\\
&0.06-0.14 & 0.10 & 0.10 & 2.5
& $-0.112\pm0.041\pm0.006$ & $-0.011\pm0.055\pm0.007$
& $0.013\pm0.076\pm0.011$ & $0.002\pm0.077\pm0.009$\\
&0.14-0.30 & 0.20 & 0.11 & 2.9
& $-0.045\pm0.047\pm0.006$ & $-0.015\pm0.063\pm0.005$
& $0.052\pm0.090\pm0.034$ & $0.078\pm0.089\pm0.009$\\
&0.30-0.70 & 0.42 & 0.12 & 3.5
& $0.060\pm0.074\pm0.014$ & $0.200\pm0.099\pm0.010$
& $0.136\pm0.147\pm0.068$ & $0.143\pm0.139\pm0.005$\\
\noalign{\smallskip}
\hline\noalign{\smallskip}
\multirow{4}{*}{\rotatebox{90}{\mbox{$x_N$}}}
&0.03-0.07 & 0.11 & 0.05 & 1.4
& $-0.053\pm0.038\pm0.002$ & $0.008\pm0.051\pm0.003$
& $0.062\pm0.074\pm0.003$ & $0.064\pm0.070\pm0.003$\\
&0.07-0.10 & 0.11 & 0.08 & 2.1
& $0.006\pm0.041\pm0.004$ & $-0.011\pm0.056\pm0.007$
& $0.108\pm0.078\pm0.014$ & $-0.085\pm0.079\pm0.005$\\
&0.10-0.15 & 0.14 & 0.12 & 3.1
& $-0.011\pm0.047\pm0.004$ & $0.043\pm0.064\pm0.014$
& $-0.004\pm0.090\pm0.021$ & $-0.112\pm0.088\pm0.009$\\
&0.15-0.35 & 0.20 & 0.20 & 5.0
& $-0.142\pm0.066\pm0.011$ & $-0.003\pm0.091\pm0.024$
& $0.199\pm0.128\pm0.034$ & $0.065\pm0.126\pm0.017$\\
\noalign{\smallskip}
\hline\noalign{\smallskip}
\multirow{4}{*}{\rotatebox{90}{\mbox{$Q^{2}$[GeV$^2$]}}}
&1.0-1.5 & 0.09 & 0.06 & 1.2
& $-0.037\pm0.042\pm0.004$ & $-0.062\pm0.056\pm0.006$
& $0.008\pm0.078\pm0.009$ & $0.083\pm0.080\pm0.007$\\
&1.5-2.3 & 0.11 & 0.08 & 1.9
& $-0.006\pm0.041\pm0.006$ & $0.054\pm0.055\pm0.005$
& $0.047\pm0.079\pm0.010$ & $-0.150\pm0.078\pm0.007$\\
&2.3-3.5 & 0.14 & 0.11 & 2.8
& $-0.047\pm0.046\pm0.003$ & $0.001\pm0.061\pm0.006$
& $0.103\pm0.085\pm0.007$ & $0.027\pm0.086\pm0.007$\\
&3.5-10.0 & 0.20 & 0.17 & 4.9
& $-0.069\pm0.050\pm0.005$ & $0.045\pm0.067\pm0.016$
& $0.166\pm0.095\pm0.010$ & $-0.011\pm0.095\pm0.007$\\
\noalign{\smallskip}
\hline\noalign{\smallskip}
\end{tabular}
\label{tb:table2}
\end{center}
\end{table}

\begin{table}
\caption{Results for azimuthal Fourier amplitudes of the 
single-beam-helicity charge asymmety $\CalACPM$, extracted from 
longitudinally polarized deuteron data, for $P_\ell=-0.530$ and a tensor 
polarization of $P_{zz}=0.827$. Not included is the 2.2$\%$ scale 
uncertainty for $\sin(n\phi)$ asymmetry amplitudes arising from the 
uncertainty in the measurement of the beam polarization.}
\tiny
\begin{center}
\begin{tabular}{r|ccccrrr}
\noalign{\smallskip}
\hline\noalign{\smallskip}
\multicolumn{2}{c}{Kinematic bin}
&$\langle -t \rangle$ &$\langle x_N \rangle$ &$\langle Q^2
\rangle$ &$\ACPMM^{\cos \, (0\phi)}$\hspace{0.9cm}
&$\ACPMM^{\cos \, \phi}$\hspace{0.9cm}
&$\ACPMM^{\cos \, (2\phi)}$\hspace{0.9cm} \\
\multicolumn{2}{c}{} &[GeV$^2$] & &[GeV$^2$] &$\pm \rm \delta_{stat} \pm
\rm \delta_{syst}$ &$\pm \rm \delta_{stat} \pm \rm \delta_{syst}$ &$\pm
\rm \delta_{stat} \pm \rm \delta_{syst}$ \\
\noalign{\smallskip}
\hline\noalign{\smallskip}
\multicolumn{2}{c}{Overall} & 0.13 & 0.10 & 2.5 & $-0.012\pm0.018
\pm0.034$ & $0.065\pm0.026\pm0.009$ & $0.017\pm0.026\pm0.003$\\
\noalign{\smallskip}
\hline\noalign{\smallskip}
\multirow{4}{*}{\rotatebox{90}{\mbox{$-t$[GeV$^2$]}}}
&0.00-0.06 & 0.03 & 0.08 & 1.9 & $0.006\pm0.030\pm0.031$
& $0.001\pm0.041\pm0.012$ & $0.005\pm0.042\pm0.009$\\
&0.06-0.14 & 0.10 & 0.10 & 2.5 & $0.074\pm0.035\pm0.034$
& $0.037\pm0.049\pm0.008$ & $0.046\pm0.049\pm0.006$\\
&0.14-0.30 & 0.20 & 0.11 & 2.9 & $-0.098\pm0.036\pm0.031$
& $0.139\pm0.052\pm0.008$ & $-0.007\pm0.051\pm0.005$\\
&0.30-0.70 & 0.42 & 0.12 & 3.5 & $-0.086\pm0.058\pm0.028$
& $0.159\pm0.088\pm0.007$ & $0.056\pm0.079\pm0.008$\\
\noalign{\smallskip}
\hline\noalign{\smallskip}
\multirow{4}{*}{\rotatebox{90}{\mbox{$x_N$}}}
&0.03-0.07 & 0.11 & 0.05 & 1.4 & $-0.046\pm0.031\pm0.035$
& $0.042\pm0.044\pm0.009$ & $-0.078\pm0.043\pm0.005$\\
&0.07-0.10 & 0.11 & 0.08 & 2.1 & $0.025\pm0.035\pm0.030$
& $0.005\pm0.050\pm0.005$ & $0.071\pm0.048\pm0.003$\\
&0.10-0.15 & 0.14 & 0.12 & 3.1 & $0.007\pm0.040\pm0.028$
& $0.038\pm0.055\pm0.010$ & $0.034\pm0.055\pm0.004$\\
&0.15-0.35 & 0.20 & 0.20 & 5.0 & $-0.052\pm0.054\pm0.024$
& $0.194\pm0.077\pm0.018$ & $0.091\pm0.079\pm0.004$\\
\noalign{\smallskip}
\hline\noalign{\smallskip}
\multirow{4}{*}{\rotatebox{90}{\mbox{$Q^{2}$[GeV$^2$]}}}
&1.0-1.5 & 0.09 & 0.06 & 1.2 & $-0.059\pm0.034\pm0.039$
& $0.119\pm0.047\pm0.007$ & $0.019\pm0.049\pm0.006$\\
&1.5-2.3 & 0.11 & 0.08 & 1.9 & $0.002\pm0.034\pm0.034$
& $0.004\pm0.049\pm0.004$ & $-0.023\pm0.047\pm0.002$\\
&2.3-3.5 & 0.14 & 0.11 & 2.8 & $0.012\pm0.038\pm0.028$
& $0.034\pm0.053\pm0.005$ & $0.087\pm0.054\pm0.004$\\
&3.5-10.0 & 0.20 & 0.17 & 4.9 & $-0.005\pm0.041\pm0.022$
& $0.111\pm0.058\pm0.010$ & $-0.004\pm0.058\pm0.005$\\
\noalign{\smallskip}
\hline\noalign{\smallskip}
\end{tabular}
\end{center}

\begin{center}
\begin{tabular}{r|ccccrrr}
\hline\noalign{\smallskip}
\multicolumn{2}{c}{Kinematic bin}
&$\langle -t \rangle$ &$\langle x_N \rangle$ &$\langle Q^2
\rangle$ &$\ACPMM^{\cos \, (3\phi)}$\hspace{0.9cm}
&$\ACPMM^{\sin \, \phi}$\hspace{0.9cm}
&$\ACPMM^{\sin \, (2\phi)}$\hspace{0.9cm} \\
\multicolumn{2}{c}{} &[GeV$^2$] & &[GeV$^2$] &$\pm \rm \delta_{stat} \pm
\rm \delta_{syst}$ &$\pm \rm \delta_{stat} \pm \rm \delta_{syst}$ &$\pm
\rm \delta_{stat} \pm \rm \delta_{syst}$ \\
\noalign{\smallskip}
\hline\noalign{\smallskip}
\multicolumn{2}{c}{Overall} & 0.13 & 0.10 & 2.5 & $0.044\pm0.026
\pm0.003$ & $-0.123\pm0.049\pm0.057$ & $0.036\pm0.049\pm0.020$\\
\noalign{\smallskip}
\hline\noalign{\smallskip}
\multirow{4}{*}{\rotatebox{90}{\mbox{$-t$[GeV$^2$]}}}
&0.00-0.06 & 0.03 & 0.08 & 1.9 & $-0.018\pm0.042\pm0.004$
& $-0.158\pm0.081\pm0.050$ & $0.109\pm0.080\pm0.005$\\
&0.06-0.14 & 0.10 & 0.10 & 2.5 & $0.075\pm0.049\pm0.004$
& $-0.156\pm0.095\pm0.036$ & $-0.021\pm0.094\pm0.015$\\
&0.14-0.30 & 0.20 & 0.11 & 2.9 & $0.053\pm0.052\pm0.005$
& $-0.126\pm0.098\pm0.021$ & $0.018\pm0.099\pm0.011$\\
&0.30-0.70 & 0.42 & 0.12 & 3.5 & $0.098\pm0.077\pm0.007$
& $0.141\pm0.142\pm0.029$ & $-0.015\pm0.153\pm0.015$\\
\noalign{\smallskip}
\hline\noalign{\smallskip}
\multirow{4}{*}{\rotatebox{90}{\mbox{$x_N$}}}
&0.03-0.07 & 0.11 & 0.05 & 1.4 & $-0.002\pm0.041\pm0.002$
& $-0.109\pm0.080\pm0.060$ & $-0.028\pm0.081\pm0.039$\\
&0.07-0.10 & 0.11 & 0.08 & 2.1 & $0.028\pm0.049\pm0.003$
& $-0.069\pm0.092\pm0.059$ & $0.095\pm0.094\pm0.018$\\
&0.10-0.15 & 0.14 & 0.12 & 3.1 & $0.091\pm0.056\pm0.006$
& $-0.288\pm0.107\pm0.058$ & $0.119\pm0.107\pm0.018$\\
&0.15-0.35 & 0.20 & 0.20 & 5.0 & $0.089\pm0.075\pm0.005$
& $-0.032\pm0.147\pm0.054$ & $-0.020\pm0.141\pm0.011$\\
\noalign{\smallskip}
\hline\noalign{\smallskip}
\multirow{4}{*}{\rotatebox{90}{\mbox{$Q^{2}$[GeV$^2$]}}}
&1.0-1.5 & 0.09 & 0.06 & 1.2 & $-0.020\pm0.048\pm0.002$
& $0.020\pm0.095\pm0.041$ & $0.011\pm0.092\pm0.056$\\
&1.5-2.3 & 0.11 & 0.08 & 1.9 & $0.074\pm0.047\pm0.003$
& $-0.266\pm0.087\pm0.043$ & $0.095\pm0.092\pm0.038$\\
&2.3-3.5 & 0.14 & 0.11 & 2.8 & $0.076\pm0.053\pm0.006$
& $-0.076\pm0.101\pm0.053$ & $-0.023\pm0.099\pm0.016$\\
&3.5-10.0 & 0.20 & 0.17 & 4.9 & $0.059\pm0.058\pm0.003$
& $-0.145\pm0.112\pm0.044$ & $0.042\pm0.112\pm0.011$\\
\noalign{\smallskip}
\hline\noalign{\smallskip}
\end{tabular}
\label{tb:table4}
\end{center}
\end{table}

\begin{table}
\caption{Results for azimuthal Fourier amplitudes of the 
single-beam-helicity charge-averaged asymmetry $\CalAP$, extracted from 
longitudinally polarized deuteron data, for $P_\ell=-0.530$ and a tensor 
polarization of $P_{zz}=0.827$. Not included is the 5.3$\%$ (5.7$\%$) 
scale uncertainty for the $\sin(n\phi)$ ($\cos(n\phi)$) asymmetry 
amplitudes arising from the uncertainties in the measurement of the target 
(beam and target) polarizations.}
\tiny
\begin{center}
\begin{tabular}{r|ccccrrrr}
\noalign{\smallskip}
\hline\noalign{\smallskip}
\multicolumn{2}{c}{Kinematic bin}
&$\langle -t \rangle$ &$\langle x_N \rangle$ &$\langle Q^2
\rangle$ &$\AP^{\cos \, (0\phi)}$\hspace{0.9cm}
&$\AP^{\cos \, \phi}$\hspace{0.9cm}
&$\AP^{\sin \, \phi}$\hspace{0.9cm}
&$\AP^{\sin \, (2\phi)}$\hspace{0.9cm} \\
\multicolumn{2}{c}{} &[GeV$^2$] & &[GeV$^2$]
&$\pm \rm \delta_{stat} \pm \rm \delta_{syst}$
&$\pm \rm \delta_{stat} \pm \rm \delta_{syst}$
&$\pm \rm \delta_{stat} \pm \rm \delta_{syst}$
&$\pm \rm \delta_{stat} \pm \rm \delta_{syst}$ \\
\noalign{\smallskip}
\hline\noalign{\smallskip}
\multicolumn{2}{c}{Overall} & 0.13 & 0.10 & 2.5
& $0.021\pm0.044\pm0.009$ & $-0.041\pm0.062 \pm0.010$
& $0.005\pm0.033\pm0.003$ & $-0.036\pm0.033\pm0.003$ \\
\noalign{\smallskip}
\hline\noalign{\smallskip}
\multirow{4}{*}{\rotatebox{90}{\mbox{$-t$[GeV$^2$]}}}
&0.00-0.06 & 0.03 & 0.08 & 1.9
& $-0.009\pm0.072\pm0.008$ & $0.087\pm0.101\pm0.011$
& $0.011\pm0.054\pm0.004$ & $0.033\pm0.054\pm0.005$ \\
&0.06-0.14 & 0.10 & 0.10 & 2.5
& $0.039\pm0.083\pm0.012$ & $-0.005\pm0.115\pm0.011$
& $0.003\pm0.063\pm0.006$ & $-0.108\pm0.062\pm0.006$ \\
&0.14-0.30 & 0.20 & 0.11 & 2.9
& $0.030\pm0.091\pm0.008$ & $-0.282\pm0.128\pm0.024$
& $0.058\pm0.068\pm0.003$ & $-0.034\pm0.068\pm0.005$ \\
&0.30-0.70 & 0.42 & 0.12 & 3.5
& $0.024\pm0.142\pm0.014$ & $-0.056\pm0.206\pm0.059$
& $-0.137\pm0.100\pm0.008$ & $-0.093\pm0.111\pm0.011$ \\
\noalign{\smallskip}
\hline\noalign{\smallskip}
\multirow{4}{*}{\rotatebox{90}{\mbox{$x_N$}}}
&0.03-0.07 & 0.11 & 0.05 & 1.4
& $0.051\pm0.075\pm0.004$ & $-0.121\pm0.106\pm0.005$
& $-0.001\pm0.054\pm0.002$ & $-0.008\pm0.056\pm0.002$ \\
&0.07-0.10 & 0.11 & 0.08 & 2.1
& $-0.014\pm0.083\pm0.012$ & $-0.049\pm0.117\pm0.021$
& $0.020\pm0.062\pm0.004$ & $-0.043\pm0.062\pm0.006$ \\
&0.10-0.15 & 0.14 & 0.12 & 3.1
& $-0.158\pm0.095\pm0.019$ & $0.033\pm0.133\pm0.030$
& $0.093\pm0.072\pm0.006$ & $-0.029\pm0.071\pm0.006$ \\
&0.15-0.35 & 0.20 & 0.20 & 5.0
& $0.228\pm0.135\pm0.027$ & $0.093\pm0.187\pm0.044$
& $-0.108\pm0.100\pm0.011$ & $-0.069\pm0.095\pm0.011$ \\
\noalign{\smallskip}
\hline\noalign{\smallskip}
\multirow{4}{*}{\rotatebox{90}{\mbox{$Q^{2}$[GeV$^2$]}}}
&1.0-1.5 & 0.09 & 0.06 & 1.2
& $0.028\pm0.085\pm0.009$ & $-0.108\pm0.116\pm0.013$
& $-0.011\pm0.065\pm0.008$ & $0.019\pm0.065\pm0.005$ \\
&1.5-2.3 & 0.11 & 0.08 & 1.9
& $-0.029\pm0.082\pm0.006$ & $-0.077\pm0.119\pm0.009$
& $-0.025\pm0.059\pm0.005$ & $-0.140\pm0.061\pm0.005$ \\
&2.3-3.5 & 0.14 & 0.11 & 2.8
& $0.018\pm0.091\pm0.012$ & $-0.049\pm0.125\pm0.013$
& $0.047\pm0.069\pm0.004$ & $-0.011\pm0.066\pm0.005$ \\
&3.5-10.0 & 0.20 & 0.17 & 4.9
& $0.078\pm0.100\pm0.022$ & $0.127\pm0.142\pm0.020$
& $0.024\pm0.075\pm0.005$ & $-0.013\pm0.075\pm0.007$ \\
\noalign{\smallskip}
\hline\noalign{\smallskip}
\end{tabular}
\label{tb:table5}
\end{center}
\end{table}

\begin{table}
\caption{Results for azimuthal Fourier amplitudes of the 
single-beam-helicity beam-charge$\otimes$target-spin asymmetry $\CalACP$, 
extracted from longitudinally polarized deuteron data, for $P_\ell=-0.530$ 
and a tensor polarization of $P_{zz}=0.827$. Not included is the 5.3$\%$ 
(5.7$\%$) scale uncertainty for the $\sin(n\phi)$ ($\cos(n\phi)$) 
asymmetry amplitudes arising from the uncertainties in the measurement of 
the target (beam and target) polarizations.}
\tiny
\begin{center}
\begin{tabular}{r|ccccrrr}
\noalign{\smallskip}
\hline\noalign{\smallskip}
\multicolumn{2}{c}{Kinematic bin}
&$\langle -t \rangle$ &$\langle x_N \rangle$ &$\langle Q^2
\rangle$ &$\ACP^{\cos \, (0\phi)}$\hspace{0.9cm} &$\ACP^{\cos \,
\phi}$\hspace{0.9cm} &$\ACP^{\cos \, (2\phi)}$\hspace{0.9cm} \\
\multicolumn{2}{c}{} &[GeV$^2$] & &[GeV$^2$] &$\pm \rm \delta_{stat}
\pm \rm \delta_{syst}$ &$\pm \rm \delta_{stat} \pm
\rm \delta_{syst}$ &$\pm \rm \delta_{stat} \pm \rm \delta_{syst}$ \\
\noalign{\smallskip}
\hline\noalign{\smallskip}
\multicolumn{2}{c}{Overall} & 0.13 & 0.10 & 2.5 & $-0.082\pm0.044
\pm0.002$ & $0.148\pm0.062 \pm0.007$ & $-0.044\pm0.057\pm0.002$ \\
\noalign{\smallskip}
\hline\noalign{\smallskip}
\multirow{4}{*}{\rotatebox{90}{\mbox{$-t$[GeV$^2$]}}}
&0.00-0.06 & 0.03 & 0.08 & 1.9 & $-0.089\pm0.072\pm0.002$
& $0.124\pm0.100\pm0.006$ & $-0.161\pm0.095\pm0.003$ \\
&0.06-0.14 & 0.10 & 0.10 & 2.5 & $-0.071\pm0.082\pm0.002$
& $0.147\pm0.114\pm0.007$ & $-0.225\pm0.110\pm0.006$ \\
&0.14-0.30 & 0.20 & 0.11 & 2.9 & $-0.152\pm0.091\pm0.003$
& $0.262\pm0.127\pm0.008$ & $0.092\pm0.114\pm0.005$ \\
&0.30-0.70 & 0.42 & 0.12 & 3.5 & $0.190\pm0.142\pm0.014$
& $0.244\pm0.205\pm0.006$ & $0.547\pm0.172\pm0.014$ \\
\noalign{\smallskip}
\hline\noalign{\smallskip}
\multirow{4}{*}{\rotatebox{90}{\mbox{$x_N$}}}
&0.03-0.07 & 0.11 & 0.05 & 1.4 & $-0.044\pm0.075\pm0.003$
& $0.101\pm0.109\pm0.001$ & $0.019\pm0.093\pm0.003$ \\
&0.07-0.10 & 0.11 & 0.08 & 2.1 & $-0.170\pm0.082\pm0.004$
& $0.210\pm0.116\pm0.006$ & $-0.132\pm0.108\pm0.003$ \\
&0.10-0.15 & 0.14 & 0.12 & 3.1 & $0.058\pm0.095\pm0.002$
& $0.323\pm0.132\pm0.010$ & $-0.239\pm0.123\pm0.007$ \\
&0.15-0.35 & 0.20 & 0.20 & 5.0 & $-0.281\pm0.135\pm0.012$
& $-0.045\pm0.188\pm0.009$ & $0.269\pm0.173\pm0.012$ \\
\noalign{\smallskip}
\hline\noalign{\smallskip}
\multirow{4}{*}{\rotatebox{90}{\mbox{$Q^{2}$[GeV$^2$]}}}
&1.0-1.5 & 0.09 & 0.06 & 1.2 & $-0.100\pm0.085\pm0.002$
& $-0.016\pm0.116\pm0.001$ & $-0.032\pm0.108\pm0.005$ \\
&1.5-2.3 & 0.11 & 0.08 & 1.9 & $-0.097\pm0.082\pm0.002$
& $0.140\pm0.119\pm0.005$ & $-0.069\pm0.104\pm0.003$ \\
&2.3-3.5 & 0.14 & 0.11 & 2.8 & $-0.002\pm0.089\pm0.001$
& $0.483\pm0.123\pm0.010$ & $-0.068\pm0.120\pm0.002$ \\
&3.5-10.0 & 0.20 & 0.17 & 4.9 & $-0.145\pm0.100\pm0.003$
& $-0.003\pm0.141\pm0.008$ & $0.005\pm0.131\pm0.009$ \\
\noalign{\smallskip}
\hline\noalign{\smallskip}
\end{tabular}
\end{center}

\begin{center}
\begin{tabular}{r|ccccrrr}
\hline\noalign{\smallskip}
\multicolumn{2}{c}{Kinematic bin}
&$\langle -t \rangle$ &$\langle x_N \rangle$ &$\langle Q^2
\rangle$ &$\ACP^{\sin \, \phi}$\hspace{0.9cm}
&$\ACP^{\sin \, (2\phi)}$\hspace{0.9cm} &$\ACP^{\sin \,
(3\phi)}$\hspace{0.9cm} \\
\multicolumn{2}{c}{} &[GeV$^2$] & &[GeV$^2$] &$\pm \rm \delta_{stat} \pm
\rm \delta_{syst}$ &$\pm \rm \delta_{stat} \pm \rm \delta_{syst}$
&$\pm \rm \delta_{stat} \pm \rm \delta_{syst}$ \\
\noalign{\smallskip}
\hline\noalign{\smallskip}
\multicolumn{2}{c}{Overall} & 0.13 & 0.10 & 2.5 & $-0.023\pm0.033
\pm0.028$ & $-0.035\pm0.033\pm0.008$ & $-0.009\pm0.030\pm0.003$ \\
\noalign{\smallskip}
\hline\noalign{\smallskip}
\multirow{4}{*}{\rotatebox{90}{\mbox{$-t$[GeV$^2$]}}}
&0.00-0.06 & 0.03 & 0.08 & 1.9 & $-0.032\pm0.054\pm0.033$
& $-0.116\pm0.053\pm0.007$ & $-0.064\pm0.050\pm0.004$ \\
&0.06-0.14 & 0.10 & 0.10 & 2.5 & $0.016\pm0.062\pm0.016$
& $-0.016\pm0.062\pm0.009$ & $-0.033\pm0.059\pm0.005$ \\
&0.14-0.30 & 0.20 & 0.11 & 2.9 & $-0.045\pm0.068\pm0.010$
& $0.024\pm0.067\pm0.015$ & $0.025\pm0.060\pm0.008$ \\
&0.30-0.70 & 0.42 & 0.12 & 3.5 & $-0.001\pm0.102\pm0.005$
& $0.212\pm0.115\pm0.020$ & $0.201\pm0.099\pm0.008$ \\
\noalign{\smallskip}
\hline\noalign{\smallskip}
\multirow{4}{*}{\rotatebox{90}{\mbox{$x_N$}}}
&0.03-0.07 & 0.11 & 0.05 & 1.4 & $-0.027\pm0.054\pm0.015$
& $0.006\pm0.058\pm0.005$ & $-0.007\pm0.051\pm0.001$ \\
&0.07-0.10 & 0.11 & 0.08 & 2.1 & $-0.073\pm0.061\pm0.023$
& $-0.098\pm0.062\pm0.004$ & $0.001\pm0.057\pm0.003$ \\
&0.10-0.15 & 0.14 & 0.12 & 3.1 & $0.031\pm0.072\pm0.026$
& $-0.087\pm0.072\pm0.003$ & $0.025\pm0.065\pm0.003$ \\
&0.15-0.35 & 0.20 & 0.20 & 5.0 & $0.009\pm0.100\pm0.031$
& $-0.001\pm0.097\pm0.002$ & $-0.036\pm0.090\pm0.006$ \\
\noalign{\smallskip}
\hline\noalign{\smallskip}
\multirow{4}{*}{\rotatebox{90}{\mbox{$Q^{2}$[GeV$^2$]}}}
&1.0-1.5 & 0.09 & 0.06 & 1.2 & $-0.003\pm0.065\pm0.022$
& $0.020\pm0.065\pm0.003$ & $-0.003\pm0.058\pm0.001$ \\
&1.5-2.3 & 0.11 & 0.08 & 1.9 & $-0.064\pm0.059\pm0.026$
& $-0.041\pm0.061\pm0.004$ & $-0.017\pm0.057\pm0.005$ \\
&2.3-3.5 & 0.14 & 0.11 & 2.8 & $-0.087\pm0.068\pm0.021$
& $-0.090\pm0.065\pm0.008$ & $-0.014\pm0.062\pm0.001$ \\
&3.5-10.0 & 0.20 & 0.17 & 4.9 & $0.102\pm0.075\pm0.027$
& $-0.040\pm0.075\pm0.005$ & $-0.004\pm0.070\pm0.004$ \\
\noalign{\smallskip}
\hline\noalign{\smallskip}
\end{tabular}
\label{tb:table6}
\end{center}
\end{table}

\end{document}